\documentclass[aps,pre,twocolumn,showpacs,groupedaddress,longbibliography
]{revtex4-1}

\usepackage{amssymb}
\usepackage{graphicx}
\usepackage{color}
\usepackage{bm}

\begin{document}

\date{\today}

\author{Oleg L. Berman}
\affiliation{Physics Department, New York City College of Technology, The City University of New York, Brooklyn, NY 11201}

\author{Roman Ya. Kezerashvili}
\affiliation{Physics Department, New York City College of Technology, The City University of New York, Brooklyn, NY 11201}

\author{German V. Kolmakov}%
\affiliation{Physics Department, New York City College of Technology, The City University of New York, Brooklyn, NY 11201}

\author{Leonid M. Pomirchi}%
\affiliation{Physics Department, New York City College of Technology, The City University of New York, Brooklyn, NY 11201}

\title{Spontaneous formation and non-equilibrium dynamics of a soliton-shaped  Bose-Einstein 
condensate in a trap}

\begin{abstract}
The Bose-stimulated self-organization of a quasi-two dimensional
non-equilibrium Bose-Einstein condensate in an in-plane  potential
is proposed. We obtained the solution of the nonlinear, driven-dissipative
Gross-Pitaevskii equation for a Bose-Einstein condensate trapped in an 
external asymmetric parabolic potential  within the method of the spectral expansion. 
We found that, in sharp contrast to previous observations, 
the condensate can
spontaneously acquire a soliton-like shape for spatially homogenous pumping. This condensate soliton performs
oscillatory motion in a parabolic trap and, also, can spontaneously rotate.
Stability of the condensate soliton in the spatially asymmetric trap is
analyzed. In addition to the nonlinear dynamics of non-equilibrium
Bose-Einstein condensates of ultra-cold atoms, our findings can be applied
to the condensates of quantum well excitons and cavity polaritons in
semiconductor heterostructure, and to the condensates of photons.
\end{abstract}


\maketitle

\section{Introduction}

\label{intro} Bose-Einstein condensation (BEC) and superfluidity are
hallmarks of quantum degenerate systems composed of interacting bosons. In
the BEC state, a substantial fraction of particles at low temperatures
spontaneously occupies the single lowest-energy quantum state thus, forming
a condensate in the energy space. One of famous examples of BECs is a
condensation of ultracold alkali-metal atoms at nK range of temperatures \cite%
{Anderson:95,Davis:95}. The advances ultra-low-temperature techniques have
already lead to the development of various technologies including
international atomic time keeping, the base of the Global Positioning
System, which is familiar to everyone.

Recently, significant progress in solid state physics and nanofabrication
has enabled to experimentally create a  class of condensed systems in
semiconductor heterostructures that demonstrate BEC \cite%
{Butov:02a,Snoke:02,Amo:09}. In this case, the Bose particles are excitons, 
\textit{i.e.}, electron-hole pairs coupled due to Coulomb attraction in
quasi-two-dimensional quantum wells (QW), or polaritons, quantum
superpositions of excitons and cavity photons (see \cite{Carusotto:13} for
extensive review). The effective mass of these particles is much smaller
than for their atomic counterpart and it varies from a the free electron
mass $m_e$ order to $\sim 10^{-4} m_e$ depending on the physical
realization. As a result, the solid-state systems undergo the BEC transition
at much higher temperatures $T_c$ than the atomic BEC: $T_c$ ranges from $%
\sim$ a few K to $\sim 40$ K \cite{Butov:94,Snoke:02,Carusotto:13}. Physics
of exciton and polariton BECs has already revealed exciting phenomena
including superfluidity \cite{Amo:09}, quantized vorticity \cite%
{Lagoudakis:08}, quantum solitons \cite{Amo:11,Sich:12}, and a
condensed-matter analogue of Dirac monopole \cite{Hivet:12}. Solid-state BEC
is a highly developing research field due to potential applications in
quantum and optical computing \cite{Gibbs:11,Menon:10}, nonlinear
interferometry \cite{Sturm:14}, novel light sources \cite{Deng:02a}, and
atomtronics \cite{Restrepo:14}. Room temperature BEC has also recently been
observed for photons~\cite{Klaers:10,Klaers:12}. In the latter case, the
nonlinear interactions between the light quanta, which are otherwise linear,
are provided by the dye molecules introduced into the microcavity.

In this paper, we study the dynamics of a trapped quasi-two-dimensional BEC
in the presence of an external source and damping that provides general
conditions for the Bose-Einstein condensation of particles with finite lifetime. The
reduced dimensionality naturally appears in the solid-state BEC realizations
in planar cavities \cite{Butov:94,Snoke:02,Amo:09}. In trapped atomic BECs
this corresponds to the limit case where a characteristic frequency of the
trap along one direction is much higher to those in two other directions 
\cite{Dalfovo:99}. To capture the experimental conditions with spatially
asymmetric traps, the condensate dynamics was considered for elliptic traps
where the trapping potential strengths in two orthogonal directions are not
the same. 
In our studies, we perform the simulations taking an exciton condensate as a
relevant example.

We report that, in sharp contrast to previous observations, under certain
conditions the condensate spontaneously self-localizes in a form of a 
\textit{solitary wave} with a size smaller than a typical condensate cloud
size and smaller than the excitation spot size.  We found that a few types
of the soliton-like waves can form, including condensate humps and rotating
doughnut condensates (rings). Earlier, the solitons propagating on the
background of a uniform condensate has been observed in
quasi-two-dimensional solid-state systems \cite{Amo:11,Sich:12}. In those
studies the solitons were the condensate perturbations, which have been
created artificially by perturbing the condensate density by an additional
``writing'' laser beam. In our work, the condensate itself self-organizes
into a strongly nonuniform soliton-shaped state that behaves as a
``particle''. This particle travels in a trap much like a
classical particle that oscillates in an external parabolic potential. 
Self-organization of the condensate into a soliton is caused by the interplay
of the Bose-stimulated condensate formation and the nonlinear interactions 
between the Bose particles that is, by the universal factors, which are present
in any physical realization of a BEC. 
It was also found that if the eccentricity of the
elliptic parabolic trap is high, the solitary wave does not form. In the
latter case, a conventional fluctuating condensate, which fills all the
energetically accessible area in the trap, was formed \cite%
{Voros:06,Berman:12,Berman:12b}.

The paper is organized as follows. In Sec.\ \ref{sec:model} we describe the
model used in the simulations and present the method of solution of the nonlinear, 
driven-dissipative Gross-Pitaevskii equation for  a BEC trapped in an external asymmetric parabolic potential. 
In Sec.\ \ref{sec:results} we discuss our
main findings. Our conclusions follow in Sec.~\ref{sec:conclusion}.

\section{Theoretical framework}

\label{sec:model}

In this section we describe our research methodology and present the system
parameters, for which the simulations have been done. To describe the
dynamics of the dipolar exciton condensate, we utilize the
driven-dissipative Gross-Pitaevskii equation for the condensate wave
function. 
In this work, we take  excitons in coupled semiconductor quantum wells, {\it i.e.} dipolar excitons,
as a physical realization of
non-equilibrium Bose-Einstein condensate where nonlinearity plays an important 
role~\cite{Butov:94,Butov:02,Butov:02a,Snoke:02a,Negoita:99,High:12}. 

{ The Gross-Pitaevskii equation also captures the dynamics of Bose-Einstein condensates of cavity polaritons \cite{Carusotto:13} and of ultra-cold atoms \cite{Pitaevskii:03}. In the latter case, the complex terms in the right-hand side of driven-dissipative Gross-Pitaevskii equation, Eq.\ (1) below,
correspond to the slow condensate depletion due to the cloud evaporation and to the initial injection of relatively hot atoms. The effective 
interaction strength in quasi-two-dimensional atomic condensates depends on the condensate density \cite{Pitaevskii:03} in analogy with the case of dipolar exciton condensates detailed in this Section below.}

{The model formulated below  captures  the realistic details 
of pumping in the Bose-Einstein condensed systems.  Specifically, 
the particles are injected into the system at elevated energies (frequencies)
and then relax to the low-energy states due to the nonlinear 
interaction with each other. The relevant examples of such systems are 
mentioned  in the Introduction.
For example, for the dipolar exciton condensates, the excitons are created at relatively high energies
due to coupling of hot electrons and holes generated by the external laser radiation
\cite{Damen:90}.
For polaritonic condensates, the scattering into the condensate from a thermal bath 
of non-condensed polaritons occurs at the bottleneck energy scale, which significantly exceeds
 the characteristic energies of the 
 particles in the ground state \cite{Carusotto:13}. In trapped atomic condensates, 
relatively hot atoms with the energies exceeding the ground-state energy in the trapping potential 
are initially injected into the system. 
 It has already been demonstrates that in the case of the dipolar exciton condensates, 
the account for these details results in formation of condensate 
turbulence under certain conditions \cite{Berman:12,Berman:12b}. 
In this paper we show that this model also predicts formation of long-living, soliton-like coherent structures
in the condensate.
In contrast, in the conventional approach (see, e.g. Ref.~\cite{Carusotto:13} for review), 
the pump rate does not 
depend on the particle energy thus, the important details of the nonequilibrium 
dynamics of the condensates are omitted. 
}

\subsection{Driven-dissipative Gross-Pitaevskii equation for two-dimensional
condensates}

\label{sec:eqs}

At temperatures below the BEC transition temperature, the dipolar exciton
condensate is described by the mean-field wave function $\Psi =\Psi (\bm{r}%
,t)$, which depends on the two-dimensional radius vector in the QW plane $%
\bm{r}=(x,y)$ and time $t$. The time evolution of the condensate wave
functions in an external trap is captured by the Gross-Pitaevskii equation  
\begin{equation}
i\hbar {\frac{\partial \Psi }{\partial t}}=-{\frac{\hbar ^{2}}{2m_{ex}}}%
\Delta \Psi + U(\bm{r}) \Psi +g \Psi |\Psi |^{2} + i \hbar {\left(\hat{R}-{%
\frac{1 }{2 \tau}}\right)}\Psi .  \label{eq:gp}
\end{equation}
In Eq.\ (\ref{eq:gp}), $m_{ex}$ is the exciton mass, $\Delta$ is the
two-dimensional Laplacian operator in the QW plane. The parabolic trapping
potential for the dipolar excitons is $U(\bm{r})={\frac{1}{2}}\left(\gamma
_{x}x^{2} +\gamma _{y}y^{2}\right)$, where $\gamma _{x}$ and $\gamma _{y}$
are the potential strengths in $x$ and $y$ directions, respectively. The
last term in Eq.~(\ref{eq:gp}) describes creation of the excitons due to the
interaction with the laser radiation and  exciton decay, $\tau$ is the
exciton lifetime. The source term $\propto \hat{R} \Psi (\bm{r},t)$ reflects
the fact that the condensate particle creation rate $\sim \partial | \Psi (%
\bm{r},t)|^2 /\partial t$ due to Bose-stimulated scattering into the
condensate is proportional to the condensate density $| \Psi (\bm{r},t)|^2$.
The effective interaction strength $g$ in Eq.\ (\ref{eq:gp}) for the dipolar
exciton condensate depends on the chemical potential $\mu$ in the system 
\cite{Berman:12,Berman:12b}. In the case where the exciton cloud size is
much greater than the mean exciton separation, the interaction strength is $%
g = 2\pi (e^4 D^4 \mu / \epsilon^2)^{1/3}$,  where $e$ is the electron
charge, $D$ is the interwell distance, $\epsilon$ is the dielectric constant
of the material in the gap between two quantum wells \cite%
{Berman:12,Berman:12b}.

To study the exciton condensate dynamics, we numerically integrate Eq.\ (\ref%
{eq:gp}). There are numbers of approaches for  the numerical integration of
Gross-Pitaevskii-type equations. One of the approaches is in solving Eq. (%
\ref{eq:gp}) in $\mathbf{r}$-space by discretizing it using a
Crank-Nicholson finite difference scheme \cite%
{Ruprecht:95,Adhikari:00,Adhikari:02}. In Refs.\ \onlinecite{Cerimele:00}
and \onlinecite{Bao:03} the ground-state wave function of a trapped BEC was
found by the direct minimization of the energy functional for the
Gross-Pitaevskii equation. The convergence of various methods has been
studied in Refs.\ \onlinecite{Sanz:84} and \onlinecite{Sanz:86}. The
stability and time-evolution of solutions of the driven-dissipative
multidimensional Gross-Pitaevskii equation has recently been analyzed
numerically in Ref.\ \onlinecite{Moxley:15}. The spectral representation of
the Gross-Pitaevskii equation has been considered for  atomic BEC
condensates without a trap \cite{Dyachenko:96}. In that case, the condensate
wave function was expanded using the plane-wave basis. Comprehensive reviews
for these methods are given in Refs.~\onlinecite{Antoine:13} and~%
\onlinecite{Kolmakov:14}.

In our approach, we use the spectral representation for the condensate wave
function $\Psi (\bm{r},t)$ by expanding it in terms of basis functions of
the exactly solvable stationary eigenvalue problem and for the
time-dependent coefficients of the expansion obtain the system of the
first-order differential equations that we solve numerically.

To utilize our approach let us consider the linear Hermitian part of the
Hamiltonian of Eq.~(\ref{eq:gp}) 
\begin{equation}
\hat{H}_{0}=-{\frac{\hbar ^{2}}{2m_{ex}}}\left( {\frac{\partial ^{2}}{%
\partial x^{2}}}+{\frac{\partial ^{2}}{\partial y^{2}}}\right) +{\frac{1}{2}}%
\left( \gamma _{x}x^{2}+\gamma _{y}y^{2}\right) ,
\end{equation}%
which is the Hamiltonian of a two-dimensional asymmetric harmonic oscillator. 
This Hamiltonian
enables the variable separation and factorization of the wave functions \cite{Messiah:61}, 
\begin{equation}
\hat{H}_{0}\psi _{n_{x}}(x)\psi _{n_{y}}(y)=E_{n_{x}n_{y}}\psi
_{n_{x}}(x)\psi _{n_{y}}(y),
\end{equation}
where $E_{n_{x}n_{y}}=E_{n_{x}}^{\lambda_x}+E_{n_{y}}^{\lambda_y}$, 
$E_{n_{x}}^{\lambda_x}=
\hbar \omega _{n_{x}}^{\lambda_x}$ and $E_{n_{y}}^{\lambda_y}=\hbar \omega _{n_{y}}^{\lambda_y}
$ determine the oscillator eigen frequencies $\omega _{n_{x}}^{\lambda_x}$ and $
\omega _{n_{y}}^{\lambda_y}$  in $x-$ and $y-$direction,
respectively. The functions $\psi _{n_{x}}^{\lambda_x}(x)$ and $\psi
_{n_{y}}^{\lambda_y}(y)$ are the eigenfunctions of a classical one-dimensional
harmonic oscillator problem that obey the time-independent Schr{\"{o}}dinger
equation 
\begin{equation}
-{\frac{\hbar ^{2}}{2m_{ex}}}{\frac{d^{2}\psi _{n_{\xi }}(\xi )}{%
d\xi ^{2}}}+{\frac{1}{2}}\gamma _{\xi }\xi ^{2}\psi _{n_{\xi }}(\xi
)=\hbar \omega _{n_{\xi }}\psi _{n_{\xi }}(\xi ),
\label{eq: HO}
\end{equation}%
where $\xi $ labels the $x$ and $y$ variables. The solution of Eq.~(\ref%
{eq: HO}) is \cite{Messiah:61} 
\begin{eqnarray}
\psi _{n_{x}}^{\lambda_x}(x) &=&{\frac{1}{\sqrt{\lambda _{x}}}\frac{1}{\sqrt{\pi
^{1/2}2^{n_{x}}n_{x}!}}}e^{-\frac{1}{2}\left( \frac{x}{\lambda _{x}}\right)
^{2}}H_{n_{x}}\left( \frac{x}{\lambda _{x}}\right) \text{, \ } \\
\psi _{n_{y}}^{\lambda_y}(y) &=&{\frac{1}{\sqrt{\lambda _{y}}}\frac{1}{\sqrt{\pi
^{1/2}2^{n_{y}}n_{y}!}}}e^{-\frac{1}{2}\left( \frac{y}{\lambda _{y}}\right)
^{2}}H_{n_{y}}\left( \frac{y}{\lambda _{y}}\right) ,
\end{eqnarray}%
where $\omega _{n_{x}}^{\lambda_x}=(\gamma _{x}/m_{ex})^{1/2}(n_{x}+1/2)$, $\omega
_{n_{y}}^{\lambda_y}=(\gamma _{y}/m_{ex})^{1/2}(n_{y}+1/2)$, $n_{x}=0,1,2,\ldots $%
, $n_{y}=0,1,2,\ldots $, \ $H_{n}$ are the Hermite polynomials, and $\lambda
_{x}$ and $\lambda
_{y}$ characterize the oscillator length scales in the $x$ and $y$
directions, respectively.

To capture the presence of the trap now let us expand the
condensate wave function $\Psi (\bm{r},t)$ in terms of the eigenfunctions $%
\psi _{\bm{n}}(\bm{r})=\psi _{{n_{x}}}^{\lambda_x}(x)\psi _{{n_{y}}}^{\lambda_y}(y)$ 
\begin{equation}
\Psi (\bm{r},t)=\sum_{\bm{n}}A_{\bm{n}}(t)\psi _{\bm{n}}(\bm{r}),
\label{eq:an}
\end{equation}
 where $A_{\bm{n}}(t)$ are the time-dependent coefficients of the
expansion, and $\bm{n}=(n_{x},n_{y})$ is a two-dimensional integer index.

After the substitution of the expansion (\ref{eq:an}) into Eq.\ (\ref{eq:gp}%
) one obtains the following system of the first-order differential equations
for the coefficients $A_{\bm{n}}(t)$, 
\begin{eqnarray}
i\hbar {\frac{\partial A_{\bm{n}}(t)}{\partial t}} &=&\hbar \omega _{\bm{n}%
}A_{\bm{n}}(t)+i\hbar \sum_{\bm{m}}R_{\bm{n}\bm{m}}A_{\bm{m}}(t)-i{\frac{%
\hbar }{2\tau }}A_{\bm{n}}(t)  \nonumber \\
&+&g\sum_{\bm{m},\bm{p},\bm{q}}W_{\bm{n}\bm{m}\bm{p}\bm{q}}A_{\bm{m}}(t)A_{%
\bm{p}}(t)A_{\bm{q}}^{\ast }(t),  \label{eq:a1}
\end{eqnarray}%
where 
\[
W_{\bm{n}\bm{m}\bm{p}\bm{q}}=\lambda
_{0}^{2}w_{n_{x}m_{x}p_{x}q_{x}}w_{n_{y}m_{y}p_{y}q_{y}},
\]%
\[
w_{nmpq}=\int_{-\infty }^{\infty }\!d\xi \,f_{n}(\xi )f_{m}(\xi )f_{p}(\xi
)f_{q}(\xi ),
\]%
and 
\[
f_{n}(\xi )={\frac{1}{\sqrt{\pi ^{1/2}2^{n}n!}}}e^{-\xi ^{2}/2}H_{n}(\xi ).
\]

The matrix elements of the source operator $\hat{R}$ in Eq.\ (\ref{eq:intrep}) are 
\begin{equation}
R_{\bm{n}\bm{m}}=\int d\bm{r}\,\psi _{\bm{n}}^{\ast }(\bm{r})\hat{R}\psi _{%
\bm{m}}(\bm{r}).  \label{eq:matr}
\end{equation}%
We consider the case where the excitons are created by an external
homogenous source in a given range of energies $(\hbar \omega _{1},\hbar
\omega _{2})$, where $\omega _{1}$ and $\omega _{2}$ are the boundary
frequencies of the excitation frequency range. Thus, we assume the following
form for the matrix elements (\ref{eq:matr}), 
\begin{equation}
R_{\bm{n}\bm{m}}=\delta _{\bm{n}\bm{m}}\Delta (\omega _{\bm{n}}),
\label{eq:source}
\end{equation}%
where $\Delta (\omega _{\bm{n}})=R_{0}$ if $\omega _{1}\leq \omega _{\bm{n}%
}\leq \omega _{2}$ and $\Delta (\omega _{\bm{n}})=0$ otherwise. The $R_{0}$
constant characterizes the intensity of the exciton source. In most
simulations, it was set equal $R_{0}=0.15\omega _{0}$ where $\omega
_{0}=(\gamma _{0}/m_{ex})^{1/2}$ is the oscillatory unit of frequency. 
{ Additionally, the simulations have been performed for  
$R_0=0.25\omega_0$ to demonstrate the transition to turbulence
at elevated pump rates.}
In the simulations, we
set $\omega _{1}=5\omega _{0}$ and $\omega _{2}=7\omega _{0}$.

We consider Eqs.\ (\ref{eq:gp}) and (\ref{eq:a1}) in the interaction
representation by separating the time dependence for the linearized equation, 
$A_{\bm{n}}(t)=a_{\bm{n}}(t)e^{-i\omega _{\bm{n}}t}$, where $\omega _{\bm{n}}
$ is the eigenfrequency of the mode $\bm{n}$. In this representation, Eq.\ (%
\ref{eq:a1}) reads 
\begin{eqnarray}
i\hbar {\frac{\partial a_{\bm{n}}(t)}{\partial t}} &=&i\hbar \left( R_{\bm{n}\bm{n}%
}-{\frac{1}{2\tau }}\right) a_{\bm{n}}(t)  \nonumber \\
&+&g\!\!\!\sum_{\bm{m},\bm{p},\bm{q}}\!\!\!W_{\bm{n}\bm{m}\bm{p}\bm{q}}\,a_{%
\bm{m}}(t)a_{\bm{p}}(t)a_{\bm{q}}^{\ast }(t)e^{i\Delta \omega t},
\label{eq:intrep}
\end{eqnarray}%
where $W_{\bm{n}\bm{m}\bm{p}\bm{q}}$ are the matrix elements of the dipolar
exciton interaction, $\Delta \omega =\omega _{\bm{q}}+\omega _{\bm{n}%
}-\omega _{\bm{m}}-\omega _{\bm{p}}$ is the frequency detuning, and a star
stands for the complex conjugate.

The characteristic length scales of the linearized problem are 
\begin{equation}
\lambda _{x}=\left( {\frac{\hbar ^{2}}{m\gamma _{x}}}\right) ^{1/4},\quad
\lambda _{y}=\left( {\frac{\hbar ^{2}}{m\gamma _{y}}}\right) ^{1/4}.
\end{equation}
We assume that the trap is elongated in $x$ direction thus, $\lambda
_{x}\geq \lambda _{y}$. The characteristic trapping potential  length scale
is $\lambda_0 = \sqrt{\lambda _{x}\lambda _{y}}$ and the mean trapping
potential  strength is $\gamma_0=\sqrt{\gamma_x \gamma_y}$. To characterize
the spatial asymmetry of the trapping potential $U(\bm{r})$  we introduce
its eccentricity as follows,  
\begin{equation}
\varepsilon =\sqrt{\frac{\lambda _{x}^{2}-\lambda _{y}^{2}}{\lambda _{x}^{2}}%
}.  \label{eq:eccentr}
\end{equation}
While the dependence of the condensate dynamics is studied for different
eccentricities $\varepsilon$, the total area accessible for the exciton
cloud  $S=\pi \lambda_0^2 $ in the trap $U(\bm{r})$ is fixed  to keep the
average condensate density constant at a given particle creation rate.

The trap parameters $\gamma _{x}$ and $\gamma _{y}$, which represent the
confinement strength, are expressed through the eccentricity (\ref%
{eq:eccentr}) as 
\begin{equation}
\gamma _{x}=\gamma _{0}(1-\varepsilon ^{2}),\quad \gamma _{y}={\frac{\gamma
_{0}}{1-\varepsilon ^{2}}}. 
\end{equation}
The characteristic length scales in the $x$ and $y$ directions are 
\begin{equation}
\lambda _{x}={\frac{\lambda _{0}}{(1-\varepsilon ^{2})^{1/4}}},\quad \lambda
_{y}={\lambda _{0}(1-\varepsilon ^{2})^{1/4}}. 
\end{equation}
The eigenfrequency of the oscillatory mode $\bm{n}=(n_x,n_y)$ in a trap with
the eccentricity $\varepsilon$ is 
\begin{eqnarray}
\omega _{\bm{n}}=&&\left( {\frac{\gamma _{0}}{m_{ex}}}\right) ^{1/2}\left[ 
\sqrt{ 1-\varepsilon ^{2}}(n_{x}+1/2) \nonumber \right. \\
&& +\left. {\frac{1}{\sqrt{1-\varepsilon ^{2}}}}%
(n_{y}+1/2) \right] . 
\end{eqnarray}

The initial conditions at $t=0$ were set as a Rayleigh-Jeans-like thermal
distribution 
\begin{equation}
a_{\bm{n}}(0)=\left( {\frac{k_{B}T}{\mu _{0}+\hbar \omega _{\bm{n}}}}\right)
^{1/2}e^{i\phi _{\bm{n}}}  \label{eq:ic}
\end{equation}%
with the chemical potential $\mu _{0}=\hbar \omega _{0}$, random phases $%
\phi _{\bm{n}}$ and the temperature $T=0.1\hbar \omega _{0}/k_{B}$, where $%
k_{B}$ is the Boltzmann constant.

\subsection{Simulation parameters and the integration method}

As we stated above,  we consider dipolar excitons with spatially separated electrons and holes 
  as the main 
example of non-equilibrium Bose-Einstein condensation \cite%
{Butov:94,Butov:02,Butov:02a,Snoke:02a,Negoita:99,High:12}.  
Taking GaAs heterostructures as a relevant example of such system, 
we set the dielectric constant equal $\epsilon = 13$. The simulations have
been done for $\gamma_0 = 50$ eV/cm$^2$, the exciton mass $m_{ex}=0.22 m_e$ 
where $m_e$ is the free electron mass, the interwell distance $D=4.2$ nm,
and the exciton lifetime $\tau=100$~ns,  which are the representative
parameters for the  experiments with the dipolar excitons in GaAs coupled
QWs \cite{Negoita:99,Voros:06,High:12}.  In the simulations, we express the
spatial coordinates and time in the oscillatory units of length $\Delta x
\equiv \lambda_0 = 0.9$ $\mu$m and of time $\Delta t =
(m_{ex}/\gamma_0)^{1/2} = 1.6$ ns. The unit of frequency in the simulations
is $\omega_0 = \Delta t ^ {-1}=6.3 \times 10^8$ s$^{-1}$.

As it is detailed above, we consider the dipolar exciton condensates under
the conditions of the experiments in Refs.\ %
\onlinecite{Butov:94,Butov:02,Butov:02a,Snoke:02a,Negoita:99,High:12} where
a rarefied exciton gas with the density of $n_{2D}\sim10^{9}-10^{10}$ cm$^{-2}$ is
formed in coupled QWs with the interwell separation $D > 0.3 a_B$, where $%
a_B =\epsilon \hbar^2/2 m_r e^2 \approx 7$~nm is the two-dimensional exciton
Bohr radius and $m_r$ is the exciton reduced mass.
{ Such dilute electron-hole system ($a_Bn_{2D}^{-1/2} \ll 1$) 
can be described as a weakly-non-ideal Bose-gas of excitons \cite{Keldysh:68,Moskalenko:00}. }
As it was shown in Refs.\ 
\onlinecite{Schindler:08} and \onlinecite{Laikhtman:09}, in this regime a
gas of interacting dipolar excitons can be considered as a quantum fluid and
the probability of biexciton formation is negligibly small.

We numerically integrated Eq.\ (\ref{eq:intrep}) with the 4th order
Runge-Kutta scheme with the time step of $10^{-2} \Delta t$. To obtain the
relative accuracy better than $10^{-3}$ for the condensate wave function and
the relative accuracy of $4\times 10^{-5}$ for the total number of excitons
in the condensate, we use $N_x \times N_y = 256$ basis functions in the
expansion (\ref{eq:an}) where $0\leq n_x<N_x$, $0 \leq n_y<N_y$, and $%
N_x=N_y=16$ \cite{Gothandaraman:2011}. 

{
In the present work, we utilized Graphics Processing Unit
(GPU) NVIDIA Tesla K20m  \cite{Nvidia} to solve numerically  Eqs.\ (\ref{eq:intrep}). 
The use of GPUs enabled us to  perform the simulations within
a practically reasonable amount of time and to achieve the
required accuracy.
In the GPU implementation, we separated the real and
imaginary parts in Eqs.\ (\ref{eq:intrep}) and thus, obtained a set
of real equations. 
 The interaction amplitudes $W_{\bm{n}\bm{m}\bm{p}\bm{q}}$ and the coefficients $a_{\bm{n}}(0)$ 
 were initialized on the host (processor) and then, were transferred from the host memory to the 
 device (GPU) memory prior to the start of the integration. 
 In the GPU  implementation of the code, 
 we used a two-dimensional (2D) computational grid consisting of 2D blocks,
 which in their turn consisted of a set of threads that worked in parallel. 
 We divided the problem size, $N_x \times N_y$, 
  in such a way that each thread on GPU operated on one element of the resultant  grid that is, updated the variables  at given $(n_x , n_y)$ at the current time step. 

It follows from Eqs.\ (\ref{eq:intrep}) 
that for each $\bm{n}$, the nonlinear interaction term involves a
summation  over six one-dimensional indices.  
Given the 2D computational
grid, if each thread in a block would access the
GPU memory to read the data, the performance of the kernel would 
suffer from the memory bandwidth bottleneck in addition to
the read latency. To effectively utilize the
memory bandwidth and avoid redundant read accesses, we
performed the following optimizations.
The GPU kernel has been divided into three sub-kernels, which performed partial summations 
in the nonlinear term of Eqs.\ (\ref{eq:intrep}). The first 
sub-kernel produced the summation over the two-dimensional 
index $\bm{q}$. The output of this kernel has been stored in the global memory on the device 
and then, has been used by the
second partial summation sub-kernel to produce the partial
sum over the index $\bm{p}$. The latter partial sum was used
by the third sub-kernel to produce the final sum in the nonlinear term in Eqs.\ (\ref{eq:intrep}). 
It was found  that splitting the  summation kernel into three sub-kernels
results in lower execution time than performing the complete
summation in a single kernel. 

We used a serial code for the same parameter set to  benchmark the GPU code. 
In the serial code, the explicit complex representation of Eqs.\ (\ref{eq:intrep}) has been utilized.
We found that the simulation for a single run
required $\sim 88$ hours on a GPU and showed $\sim 10 \times$
acceleration compared to the serial version of the code. 
}

\section{Results and Discussion}

\label{sec:results}

In this section we present the results of our studies. First, we consider
spontaneous formation of the dipolar exciton condensate patterns and their
dynamics in a trapping potential. Then, we analyze the details of the
pattern rotation. Finally, we consider the effect of the asymmetry of the
trapping potential and compare our findings with the existing results.

\subsection{Formation of a soliton-like condensate}

\label{sec:form}

\begin{figure}[h]
\begin{center}
\includegraphics[width=7cm]{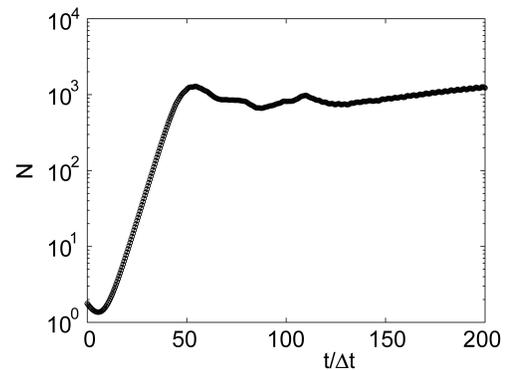}
\end{center}
\caption{Dependence of the total number of dipolar excitons in the
condensate on time in a trap with $\protect\gamma_0=50$ eV/cm$^2$ for the
trap eccentricity $\protect\varepsilon = 0.2$. The numerical unit of time $%
\Delta t$ is specified in the text.}
\label{fig:np}
\end{figure}

To study the dynamics of the dipolar exciton condensate, we determined the
condensate density distribution at different moments of time by integrating
Eq.\ (\ref{eq:gp}) with the source term (\ref{eq:source}) and initial
conditions (\ref{eq:ic}).  In our studies, we performed six independent runs
for the trap eccentricity $\varepsilon = 0.2$ and three independent runs for
all other values of the eccentricity in the range $0 \leq \varepsilon \leq
0.8$. Fig.\ \ref{fig:np} demonstrates the dependence of the total number of
dipolar excitons in the condensate as a function of time, 
\begin{equation}
N(t)=\int d^2 \bm{r} |\Psi( \bm{r},t)|^2,
\end{equation}
obtained in the simulations for the trap eccentricity $\varepsilon=0.2$. It
is seen that initially the number of excitons grows with time and then, 
saturates at $t > t_s \equiv 70 \Delta t \approx 1.1\times 10^{-7}$~s.  At $%
t \sim t_s$ the system comes to a steady state where the total number of
dipolar excitons  only slowly varies with time.

\begin{figure*}[t]
\begin{center}
\includegraphics[width=5.3cm]{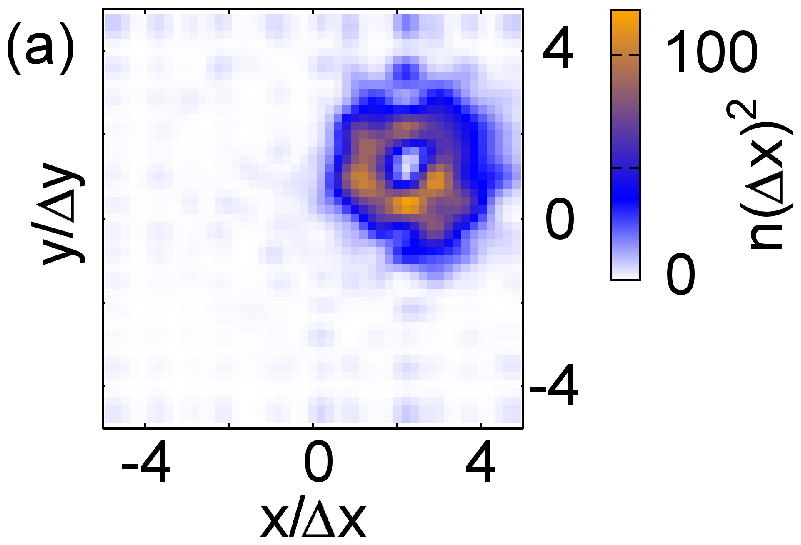} \hspace{-.5cm}
\includegraphics[width=5.3cm]{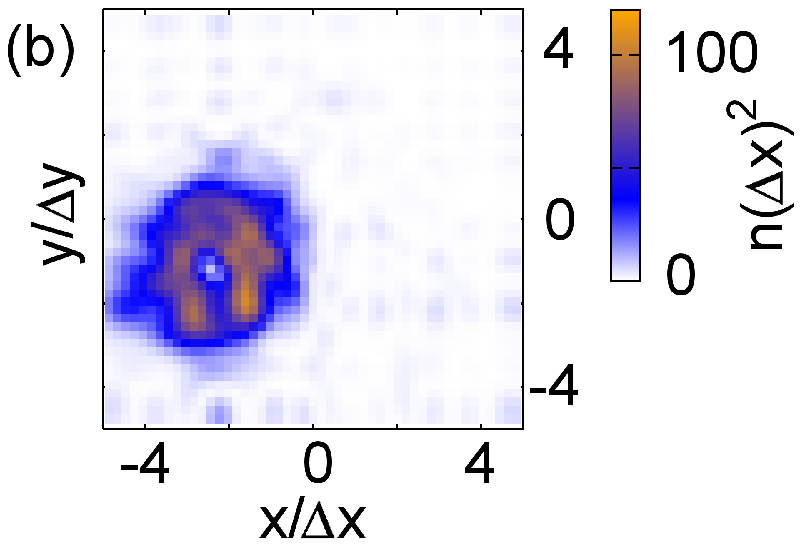} \hspace{-.5cm}
\includegraphics[width=5.3cm]{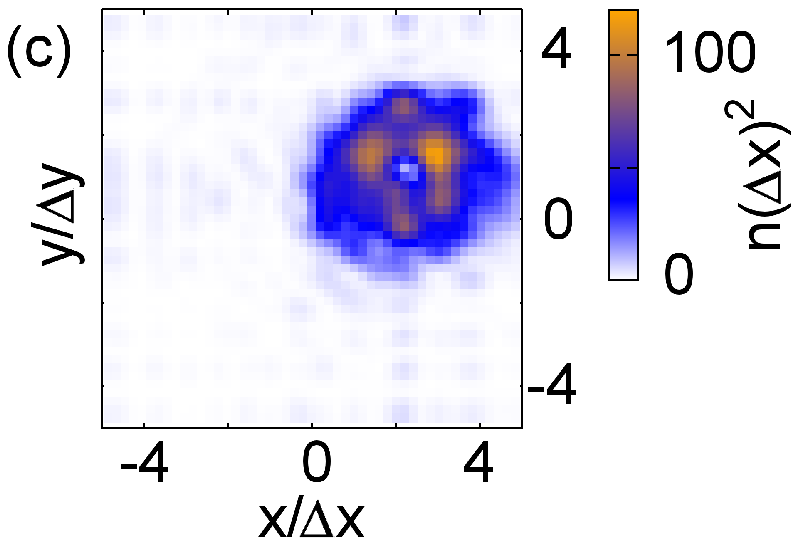}\\
\vspace{0.2cm} 
\includegraphics[width=5.3cm]{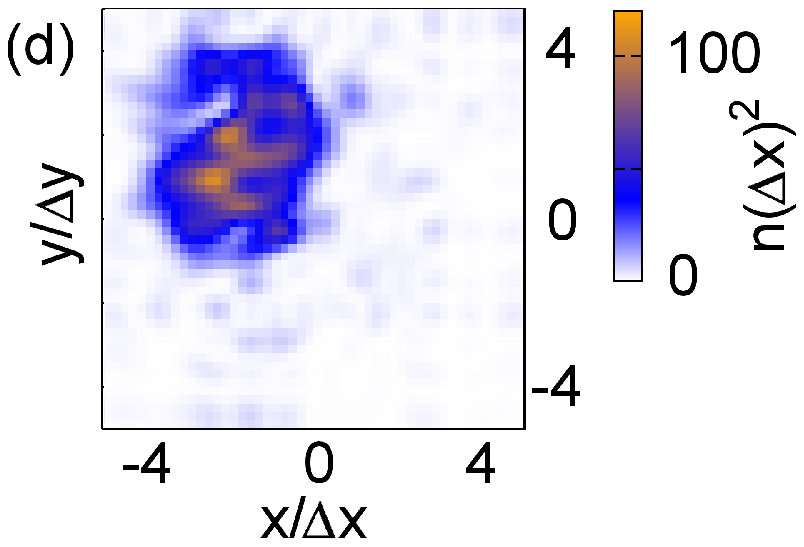} \hspace{-.5cm}
\includegraphics[width=5.3cm]{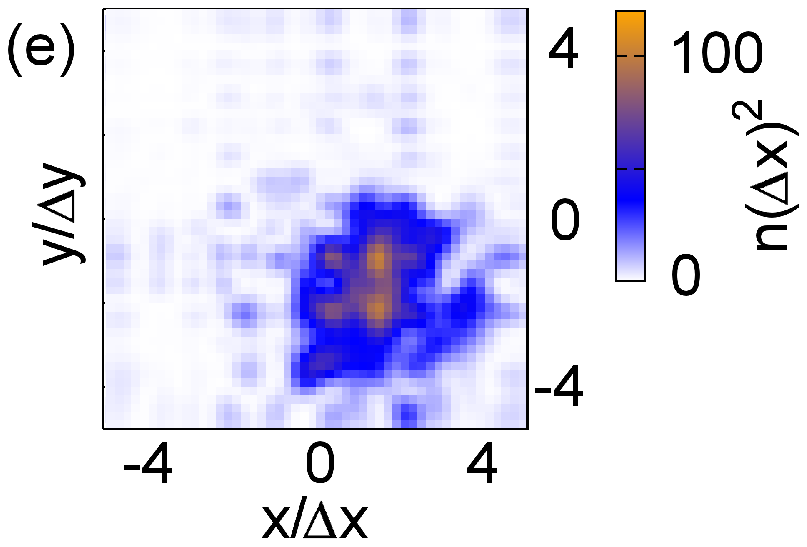} \hspace{-.5cm}
\includegraphics[width=5.3cm]{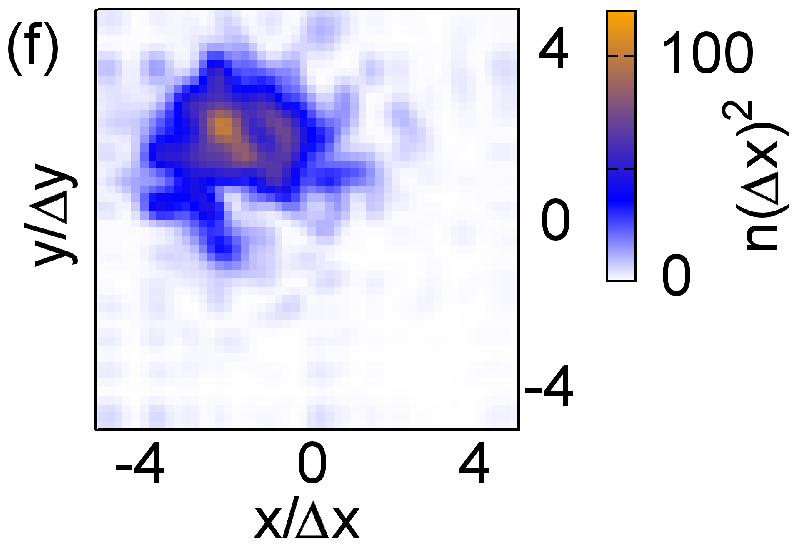}
\end{center}
\caption{(Color online) Formation of a traveling soliton-like
dipolar-exciton pattern in a trap with the eccentricity $\protect\varepsilon%
=0.2$ (upper row) and $\protect\varepsilon=0$ (lower row). The figures show
the exciton condensate density $n(\bm{r},t) = |\Psi(\bm{r},t)|^2$ in three
subsequent moments. Upper row: (a) $t = 139.5 \Delta t$, (b) $t = 142.5
\Delta t$, and (c) $t = 145.5 \Delta t$ after the external drive was turned
on. Lower row: (d) $t = 143.0 \Delta t$, (e) $t = 146.5 \Delta t$, and (f) $%
t = 149.5 \Delta t$. The vertical bars show the exciton condensate density.
The numerical unit of length $\Delta x$ is specified in the text.}
\label{fig:ring}
\end{figure*}

To characterize the dipolar exciton condensate dynamics in the steady state
at $t>t_s$, we studied the time evolution of the condensate density in the
trap. Fig.\ \ref{fig:ring} shows the graphical output from the simulations
made for the eccentricity $\varepsilon = 0.2$ (upper row) and $\varepsilon =
0$ (lower row). The results in Fig.\ \ref{fig:ring}a-c are shown for the
same data with $\varepsilon = 0.2$ as in Fig.~\ref{fig:np} above. It is seen
in Fig.\ \ref{fig:ring} that in both cases $\varepsilon \neq 0$ and $%
\varepsilon = 0$, a spatially-localized condensate pattern is formed. Below
we refer to such structures as condensate solitary waves, or condensate
solitons.

\begin{figure}[h]
\begin{center}
\includegraphics[width=7cm]{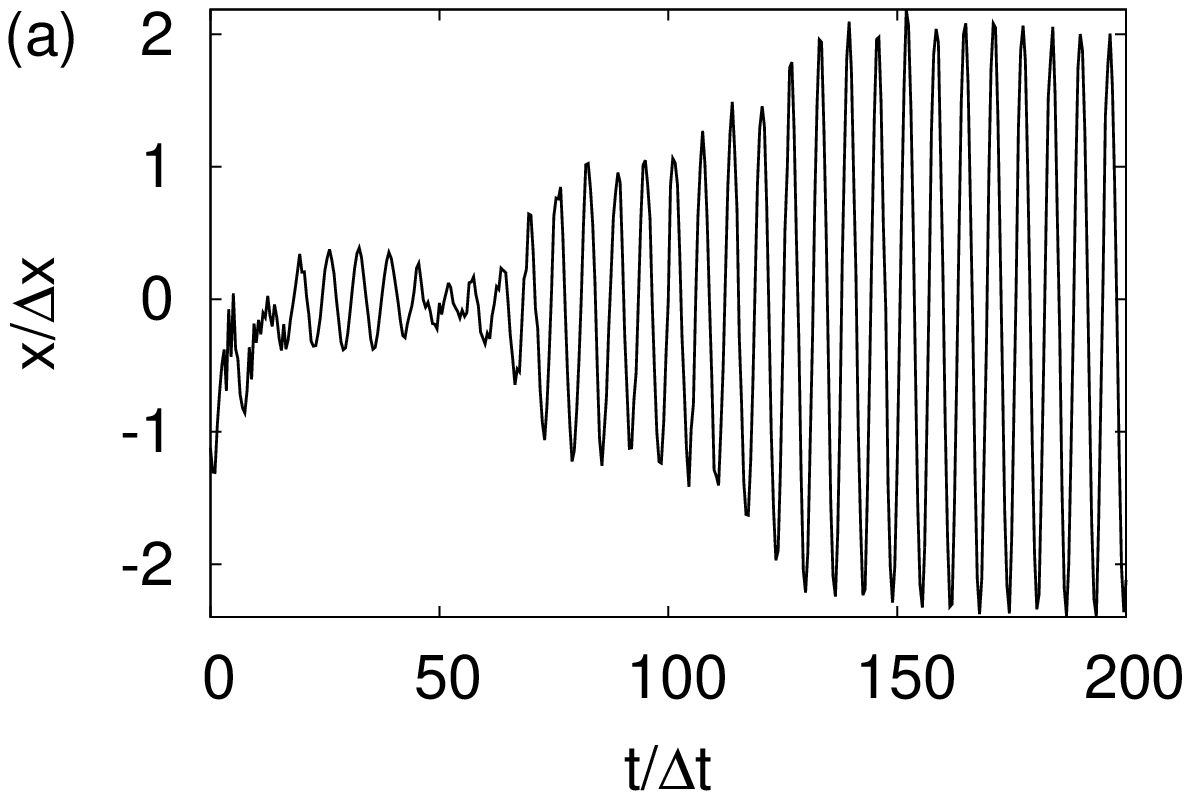}\\[0pt]
\includegraphics[width=7cm]{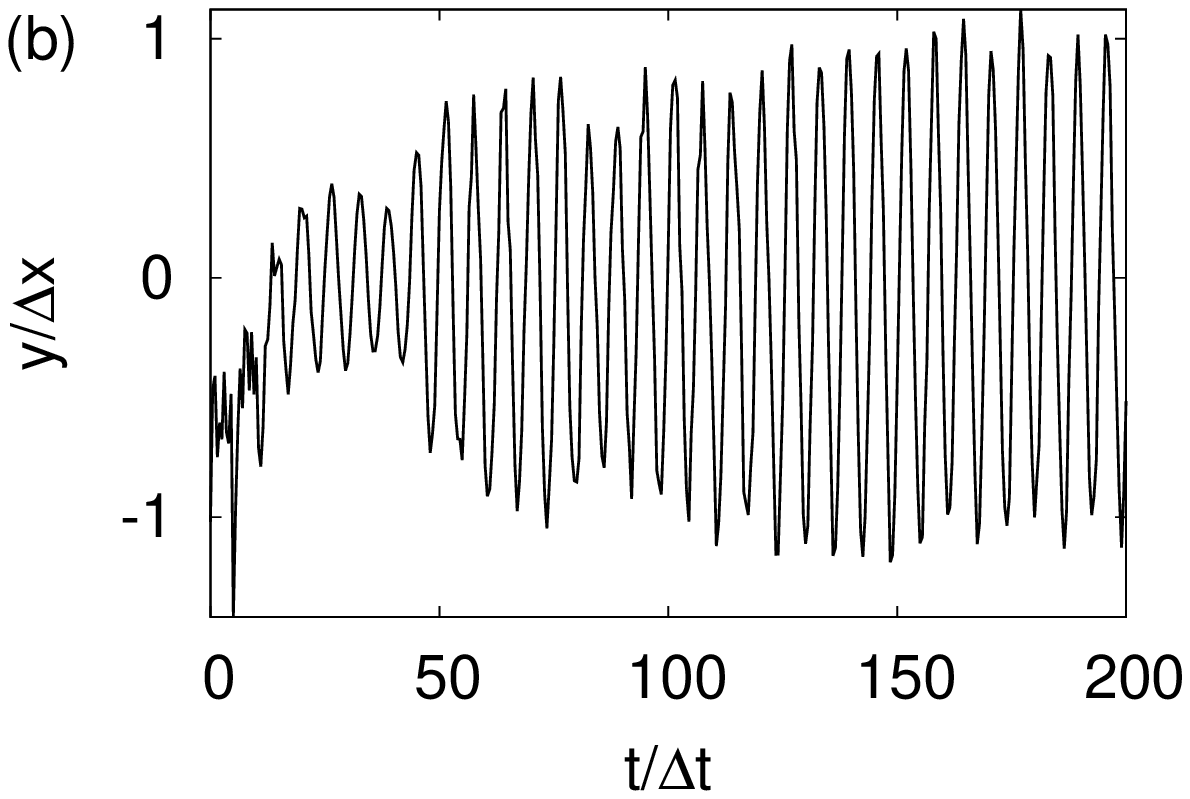}\\[0pt]
\includegraphics[width=7cm]{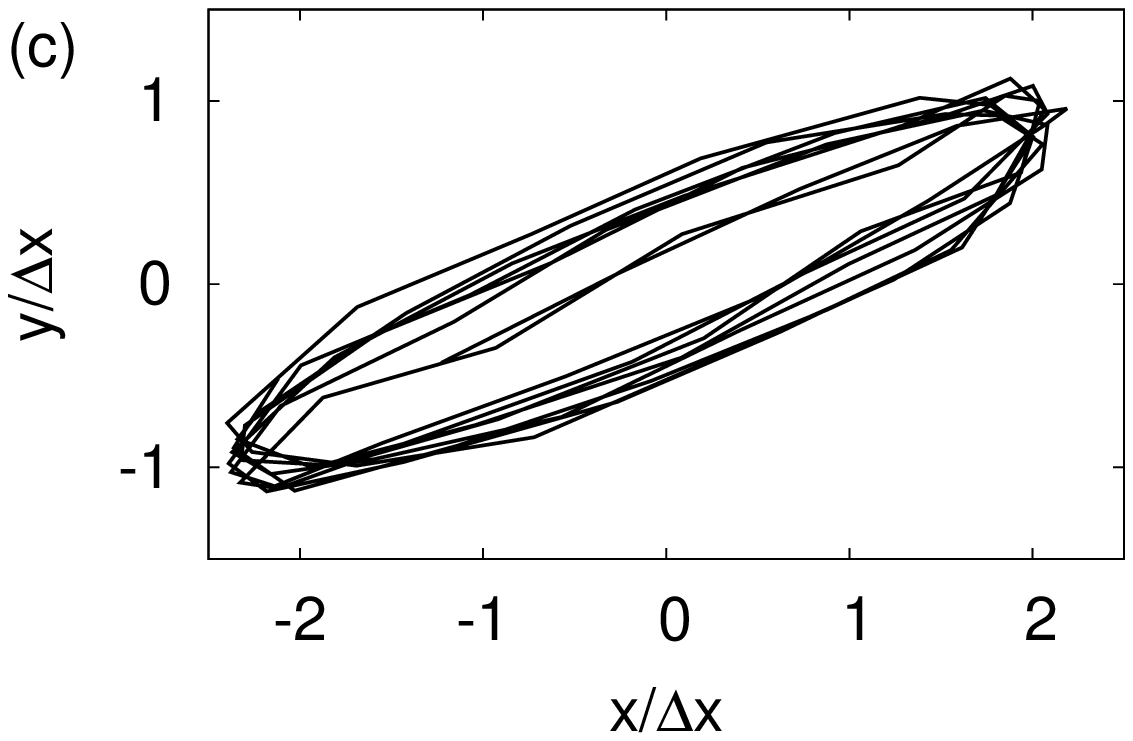}
\end{center}
\caption{Motion of (a) $x$- and (b) $y$-coordinates of the center of mass of
the condensate pattern. Figure (c) shows motion of the center of mass of the
condensate in the $(x,y)$ plane for $t\geq 150 \Delta t$, where the pattern
exhibits stable periodic oscillations in the parabolic trap. The
eccentricity of the trapping potential is $\protect\varepsilon=0.2$. The
curves in figures a-c are shown for the data in Fig.\ \protect\ref{fig:ring}%
a-c. }
\label{fig:cm}
\end{figure}

It is also seen in Fig.\ \ref{fig:ring}, the pattern travels in the trapping
potential. To characterize the condensate pattern motion, we show in Fig.\ %
\ref{fig:cm}a,b the dependence on time of the $x$ and $y$ coordinates of the
center of mass of the dipolar exciton condensate for the data presented in
Fig.\ \ref{fig:ring}a-c. The center-of-mass coordinate $\bm{r}_{cm} (t) =
(x_{cm} (t), y_{cm} (t))$ was calculated from the condensate wave function
as follows 
\begin{equation}
\bm{r}_{cm} (t) = \frac{\int d^2 \bm{r} |\Psi( \bm{r},t)|^2 \bm{r} }{\int
d^2 \bm{r} |\Psi( \bm{r},t)|^2}.  \label{eq:rcm}
\end{equation}
It is seen from Fig.\ \ref{fig:cm}a,b that the position of the center of
mass of the pattern oscillates with time. At the moment $t\sim 130 \Delta t$%
, the oscillations reach a steady state, in which their amplitude and period
only slowly varies with time. As follows from Fig.\ \ref{fig:cm}a,b, the
period of the oscillation of the pattern in the trap is $\sim 6.3 \Delta t
\approx 1.0 \times 10^{-8}$~s. The trajectory of the pattern can also be
viewed as a Lissajous figure for the condensate center of mass in a
parametric plot in the $(x_{cm}(t), y_{cm}(t))$ plane, as it is presented in
Fig.\ \ref{fig:cm}c. The distance between two farthest points of the
trajectory for the condensate center of mass estimated from Fig.\ \ref%
{fig:cm}a-c is $\sim 4.8 \Delta x \approx 4.3$ $\mu$m.

\begin{figure}[h]
\begin{center}
\includegraphics[width=7cm]{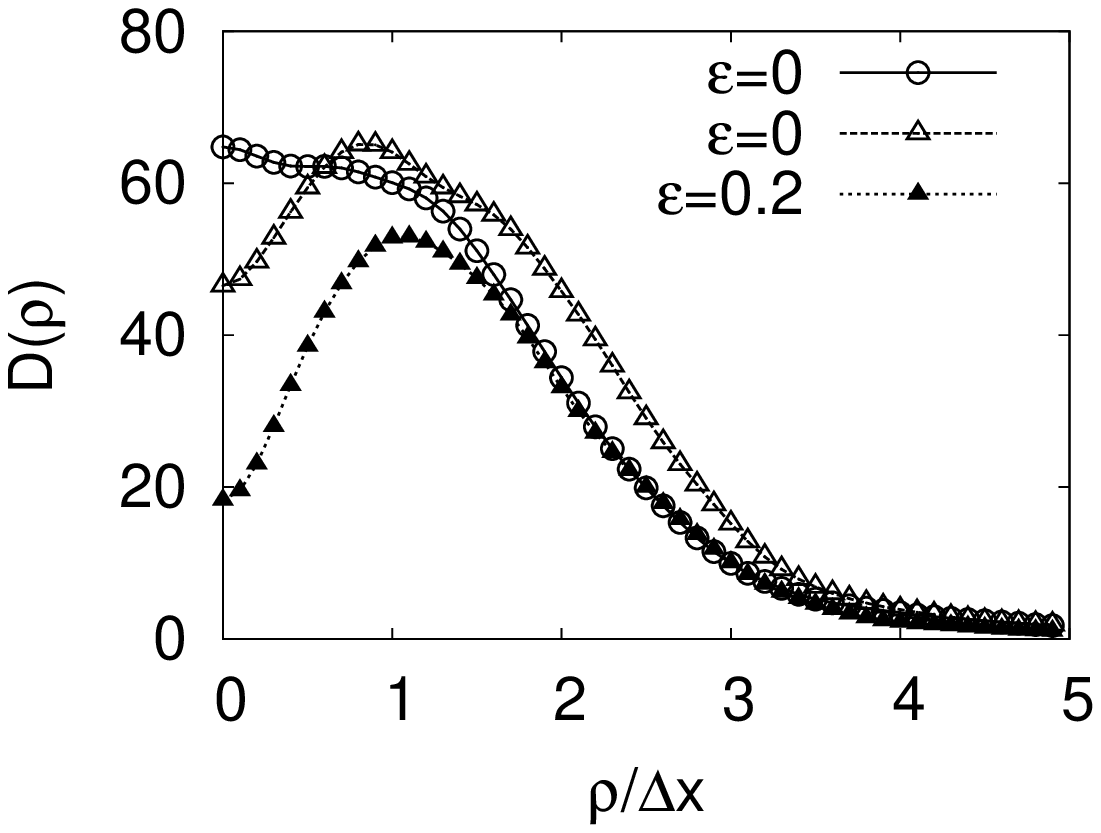}
\end{center}
\caption{Time-averaged radial density distribution (\protect\ref{eq:d}) of
the dipolar exciton condensate soliton as a function of the radial distance $%
\protect\rho$ in the center-of-mass frame for three independent runs with
the eccentricity $\protect\varepsilon = 0$ (open circles and open triangles)
and for $\protect\varepsilon=0.2$ (filled triangles). The averaging is made
for the time period between the moments $t_1 = 100 \Delta t$ and $t_2=200
\Delta t$. Data for $\protect\varepsilon = 0$ (open circles) and $\protect%
\varepsilon=0.2$ (filled triangles) are calculated for the data shown in
Fig.\ \protect\ref{fig:ring} in the upper and lower rows, respectively.
Points show the result of the calculation, curves are shown to guide the eye.
}
\label{fig:d}
\end{figure}

To characterize the internal structure of the soliton-like pattern, we
calculated the time- and angle-averaged radial density distribution function
for the condensate in the center-of-mass frame, 
\begin{equation}
D(\rho) = {\frac{1}{t_2-t_1}} \int_{t_1}^{t_2} dt \left( {\frac{1 }{2 \pi}}
\int_0^{2\pi} d\varphi_{\bm{\rho}} |\Psi( \bm{r}_{cm}(t) + \bm{\rho},t)|^2
\right).  \label{eq:d}
\end{equation}
In Eq.\ (\ref{eq:d}), $(t_2-t_1)$ is the time-averaging interval, the
integral in the round brackets is taken over the direction $\varphi_{%
\bm{\rho}} $ of the relative radius-vector $\bm{\rho}= \bm{r} - \bm{r}%
_{cm}(t)$, and we denoted $\rho = |\bm{\rho}|$. The time-averaging interval
was bound by the moments $t_1=140 \Delta t$ and $t_2=200\Delta t$. The
results of the calculations for three independent runs for $\varepsilon = 0$
and $\varepsilon=0.2$ are shown Fig.\ \ref{fig:d}. It is seen that the
characteristic radius $\rho_0$ of the pattern defined as a width at the half
height for $D(\rho)$ varies from $\sim 2 \Delta x$ to $2.5 \Delta x$. This
conclusion was validated for all independent runs performed under the same
conditions. The comparison of Figs.\ \ref{fig:cm}a-c and Fig.\ \ref{fig:d}
shows that the traveled distance for the soliton-like pattern during its
oscillation in the trap is of the order of or greater than twice the radius
of the pattern. In other words, a soliton-like pattern performs
large-amplitude oscillatory motion in a parabolic trap.

It is also seen in Fig.\ \ref{fig:d} that two types of patterns are formed.
In the pattern of the first kind, the maximum of the exciton condensate
density is positioned at its center $\bm{r}_{cm}$ (see open circles in Fig.\ %
\ref{fig:d}). This corresponds to a hump-like soliton structure. In the
pattern of the second kind, the condensate density has a minimum at the
center of the pattern (open triangles and filled triangles in Fig.\ \ref%
{fig:d}) and therefore, it corresponds to a ring-like structure.  We also
found that the structures of both kinds can be formed for the same
eccentricity (cf.\ open circles and open triangles for $\varepsilon=0$ in
Fig.\ \ref{fig:d}). It is worth noting that the runs are independent since
each run was set based on random phases in the initial conditions for the
condensate wave function at $t=0$ as described in Sec.\ \ref{sec:eqs}. The
motion of a ring structure is shown in the first row of Fig.\ \ref{fig:ring}%
, for which the minimum of the condensate density  is seen as a bright spot
at the center of the pattern, whereas a hump-like structure is shown in the
second row of the same figure.

We infer that the solitary-wave condensate formation  is caused by the
interplay of the two factors: (i) Bose-stimulated scattering of excitons
into the condensate and  (ii) the exciton-exciton repulsion in the
condensate.  The first factor results in the increased probability $\propto
n_i$ of the condensate growth in the state $i$ in the region  already
occupied by $n_i$ particles \cite{Snoke:09}.  This effect is captured by the
source term in Eq.\ (\ref{eq:gp}) as explained in Sec.\ \ref{sec:eqs}.  It
leads to the exponential-like increase of the particle occupation numbers in
the occupied states. However, if the spatial density of the condensate is
increased, the mutual repulsion of the condensate particles tends to
decrease the density and results in ``spreading'' of the particles over the
system. The nonlinear term in Eq. (\ref{eq:gp}) is responsible for this
effect. In the presence of the continuous source of the particles (\textit{%
i.e.}, external laser radiation), this is the finite lifetime of dipolar
excitons that limits the growth of the total number of the particles in the
condensate. The approach where the particles are injected into the system at
a given energy scale has earlier been used to describe the dynamics of the
Bose-Einstein condensation of ultra-cold gases \cite%
{Dyachenko:96,YuLvov:03,Zakharov:05} (see Ref.\ \onlinecite{Kolmakov:14} for
review). Specifically, in that scenario, the injected hot atoms relaxed due
to multiple collisions toward the condensate state. The condensate depletion
due to atom evaporation was captured by introducing the effective condensate
decay rate. In contrast, for the case of dipolar exciton condensate
considered in this paper,  the depletion rate is set the same for all states
to capture the fact that the dipolar exciton lifetime only weakly depends on
the exciton energy.

\begin{figure}[h]
\begin{center}
\includegraphics[width=7cm]{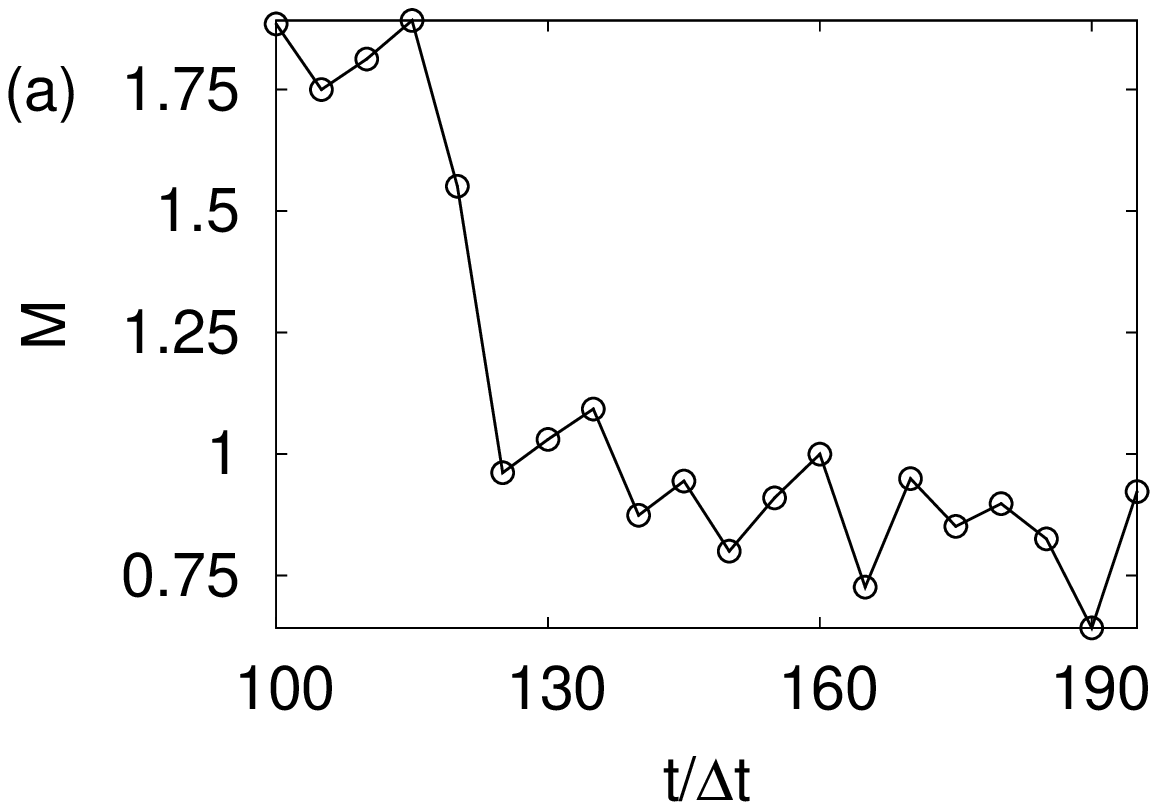}\\[0pt]
\includegraphics[width=7cm]{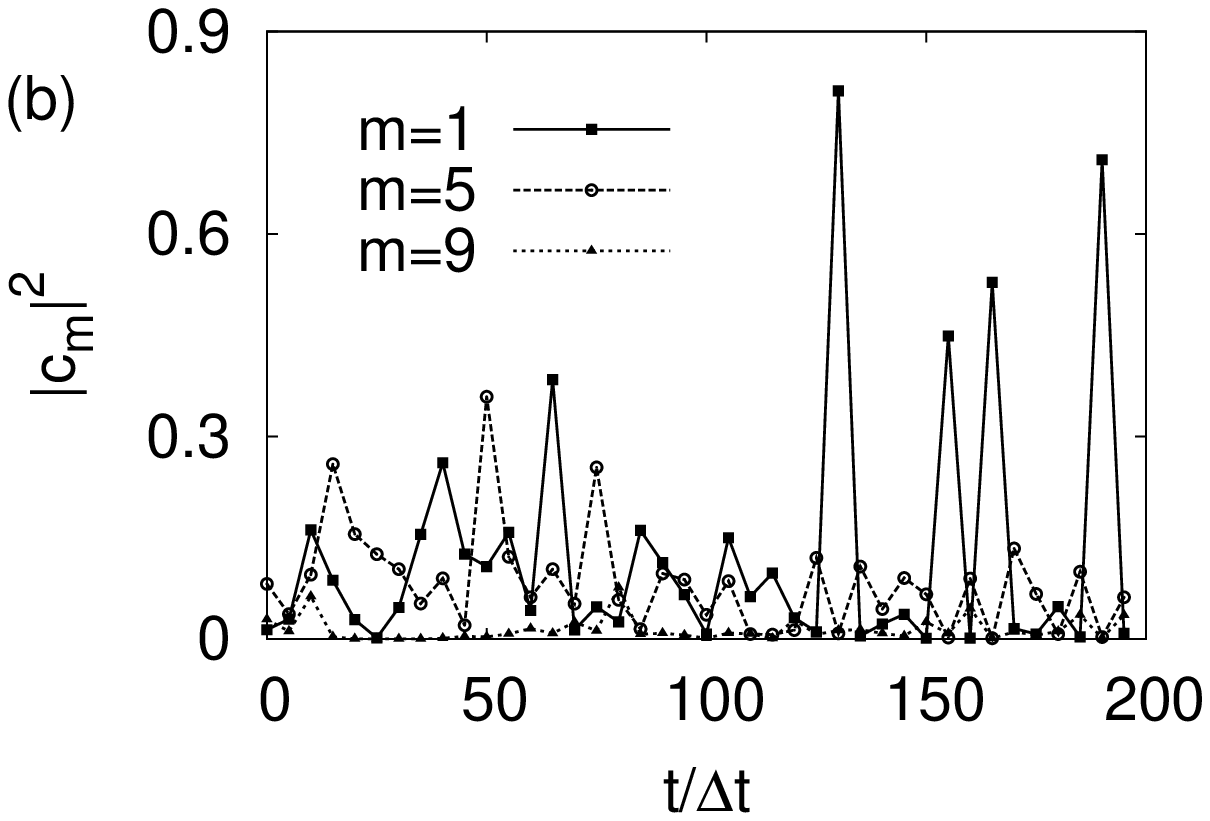}\\[0pt]
\includegraphics[width=7cm]{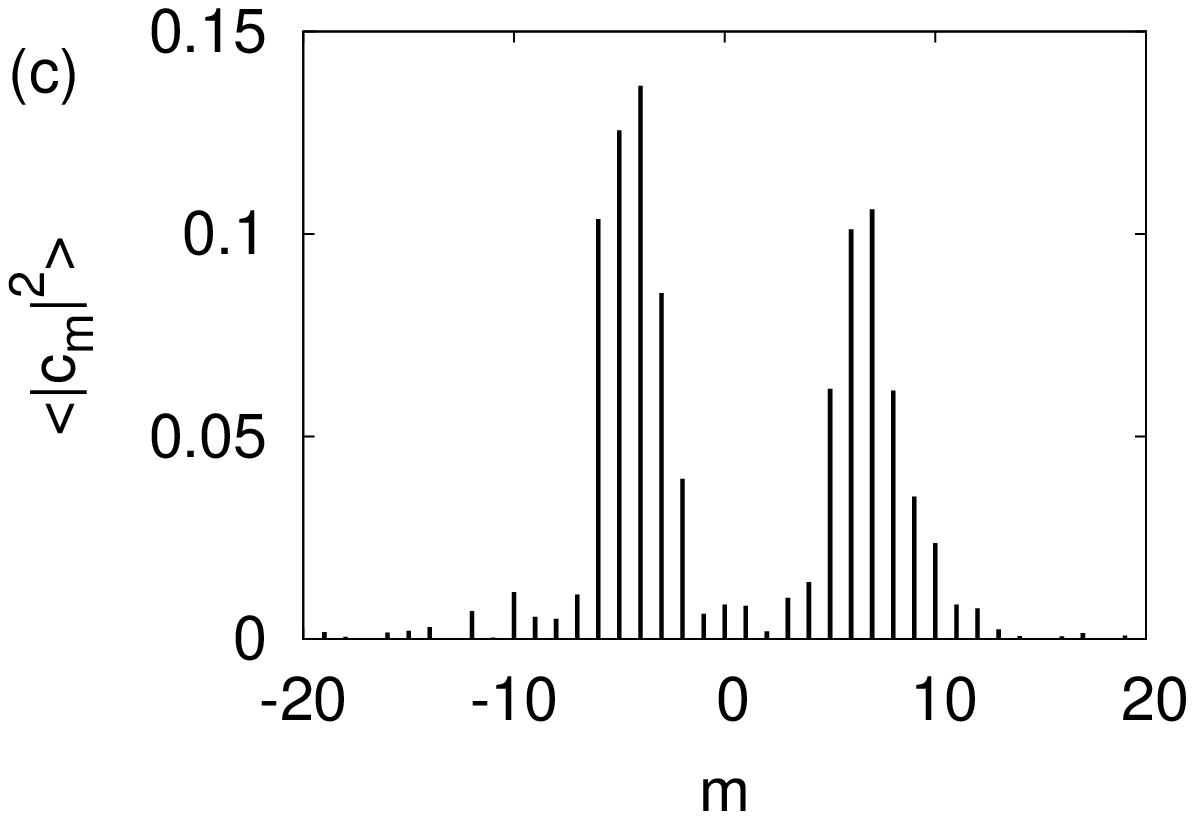}
\end{center}
\caption{ (a) Dependence of the angular momentum of the ring structure in
Fig. \protect\ref{fig:ring}a-c on time. Points show the results of the
calculation, the line segments between the points are shown to guide the
eye. The momentum is expressed in the units of $\hbar$. (b) The
time-dependence of the probabilities $|c_m(t)|^2$ of having the momentum $m$%
. (c) Time-averaged probabilities, $\langle |c_m(t)|^2\rangle$, for various
momenta $m$. The graphs in a-c are plotted for the data shown in Fig.\ 
\protect\ref{fig:ring}a-c}
\label{fig:momentum}
\end{figure}

\subsection{Ring condensate rotation}

We assume that formation of the ring structures is caused by spontaneous
rotation of the non-equilibrium condensate. It is known that for the
Bose-condensed, superfluid systems two types of rotation are possible: (i)
vortex rotation and (ii) solid-body-like rotation \cite{Khalatnikov:65}. The
vortex rotation is realized as quantized vortices in the condensate. The
latter have a velocity singularity and zero condensate density at the vortex
core \cite{Gross:61}. In addition to classical experiments with superfluid
He-II (see, \textit{e.g.}, a review in Ref.\ \onlinecite{Andronikashvili:61}%
), quantized vortices have recently been observed in Bose-Einstein
condensates of ultra-cold atoms \cite{Weiler:08} and in condensates of
exciton polaritons \cite{Lagoudakis:08,Manni:11,Keeling:08}. For the
solid-body rotation, the condensate density does not mandatory turn to zero
at the rotation center \cite{Khalatnikov:65}. Below we demonstrate that the
solid-body rotation is realized for the soliton-like patterns described
above.

To analyze the rotation of the pattern we, first, calculate the angular
momentum of the dipolar exciton condensate as a function of time 
\begin{equation}
M(t)=\int d^{2}\bm{r}\Psi ^{\ast }(\bm{r},t)\hat{M_{z}}\Psi (\bm{r},t),
\label{eq:mt}
\end{equation}%
where 
\[
\hat{\bm{M}}=[\bm{\rho}\times \hat{\bm p}]
\]%
is the angular momentum operator for the condensate in the center-of-mass
frame, $\hat{\bm p}=-i\hbar \nabla _{\bm{\rho}}\equiv -i\hbar \frac{\partial 
}{\partial \bm{\rho}}$ is the linear momentum operator, and $z$ marks the
direction perpendicular to the QW plane. 
Fig.\ \ref{fig:momentum}a demonstrates the time dependence of the angular
momentum of the ring pattern shown in Fig.\ \ref{fig:ring}a-c. We see in
Fig.~\ref{fig:momentum}a that the angular momentum of the ring pattern
fluctuates with time. We note that the angular momentum fluctuations are
natural for the dipolar exciton condensate since the system is driven by an
external pump: according to Eq.\ (\ref{eq:gp}) both the source and damping
of the excitons are present. Additionally, the ring condensate is formed in
the parabolic trap that breaks the rotational symmetry in the moving,
off-center frame of reference. Therefore, the angular momentum of the system
is not conserved. However, it is seen in Fig.\ \ref{fig:momentum}a that at
the moment of time $t\approx 140\Delta t$ the angular momentum relaxes to a
steady state and it further fluctuates about the mean value $M\approx 0.8$.

\begin{figure}[h]
\begin{center}
\includegraphics[width=7.5cm]{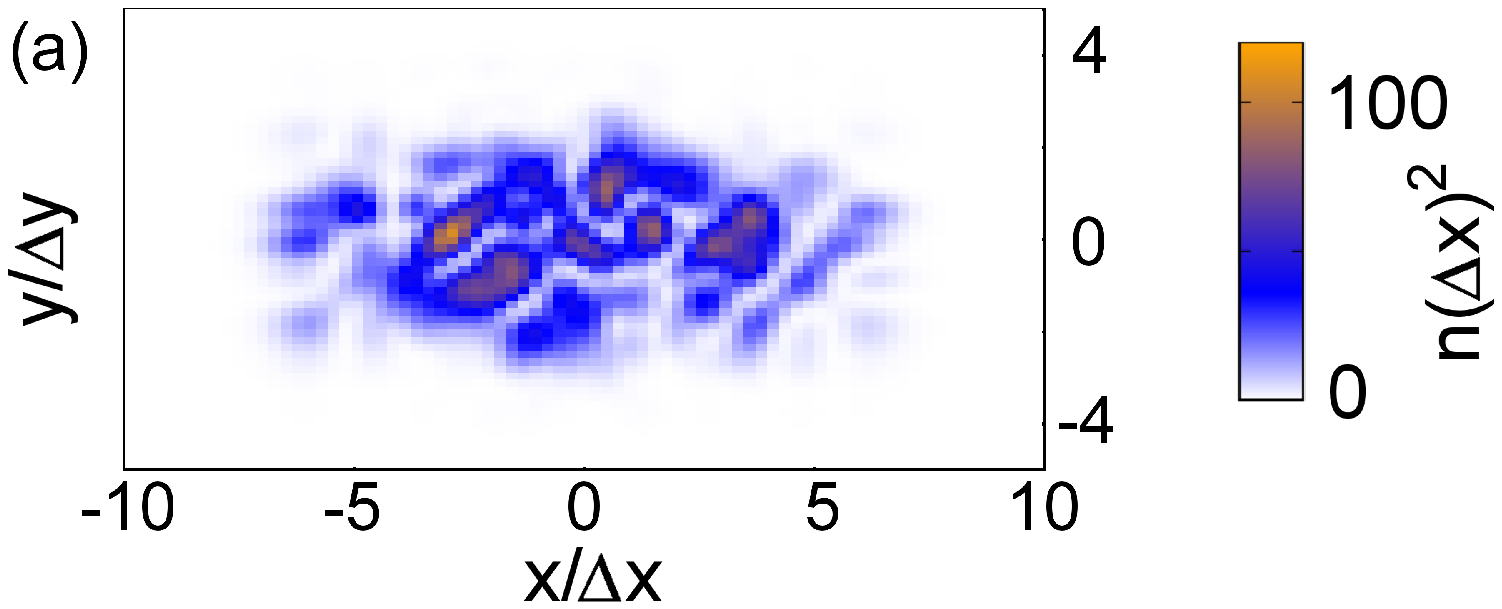}\\[0pt]
\includegraphics[width=7.5cm]{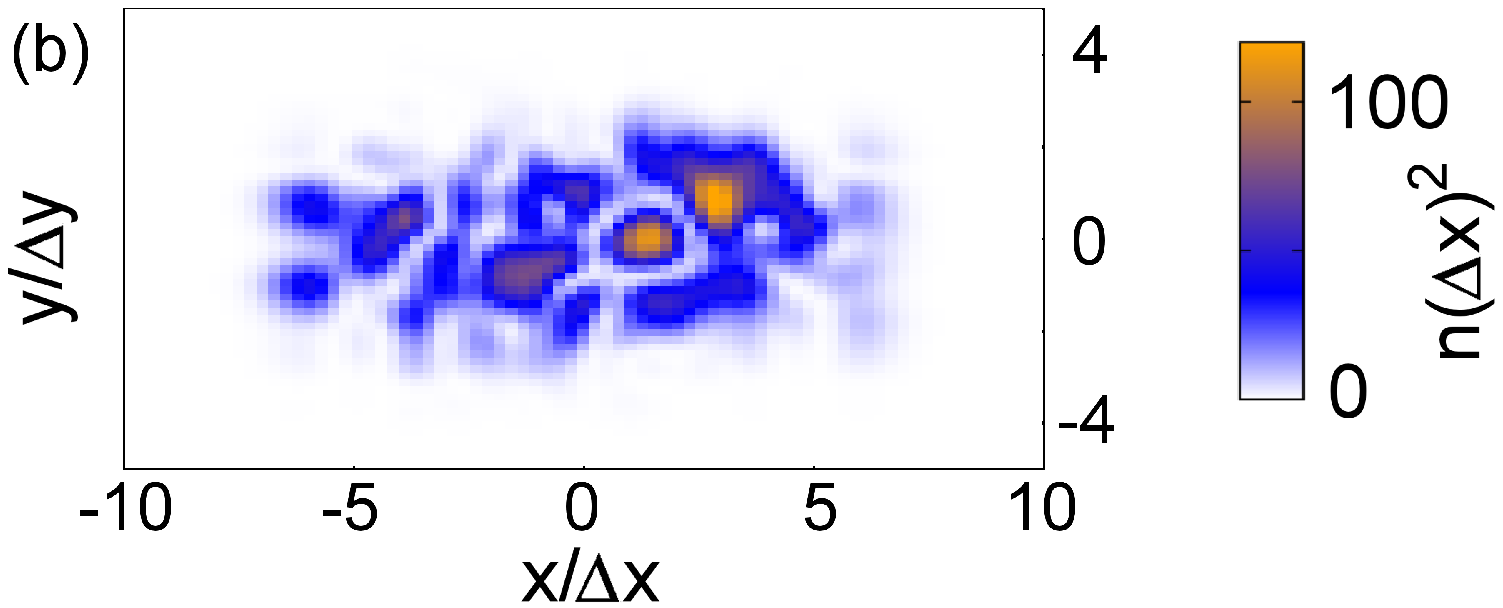}\\[0pt]
\includegraphics[width=7.5cm]{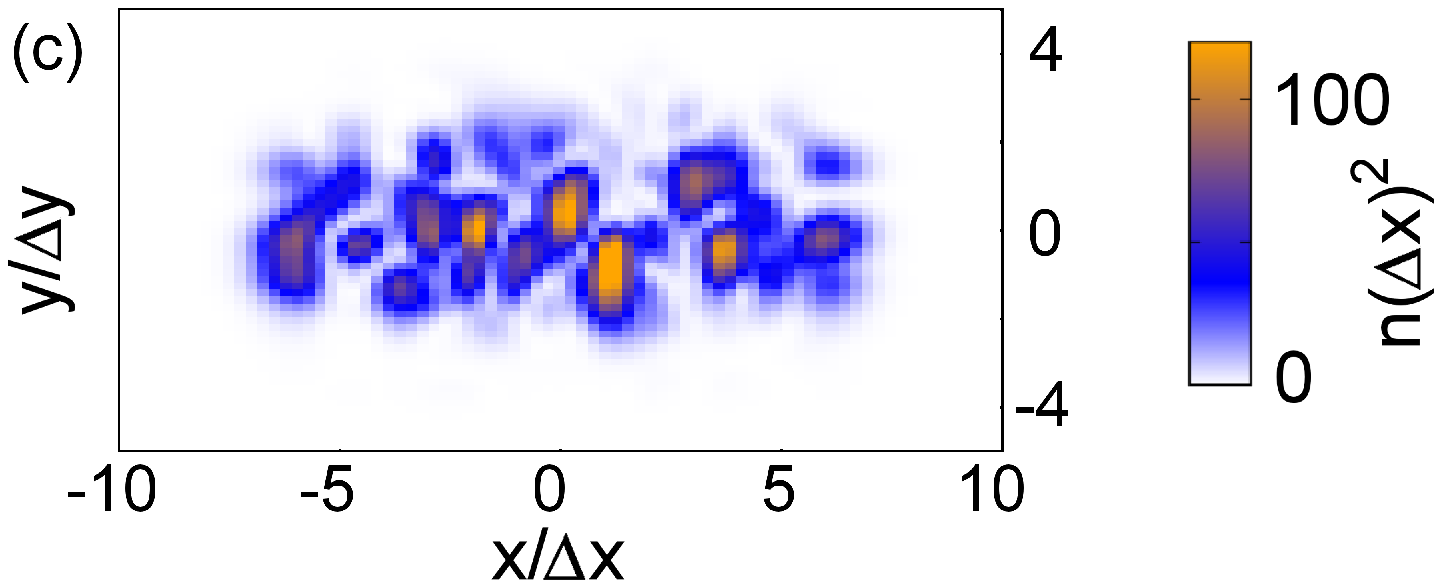}
\end{center}
\par
\caption{(Color online) Exciton density distribution in a trap with the
eccentricity $\protect\varepsilon = 0.8$ for the moments (a) $t = 139.5
\Delta t$, (b) $t = 142.5 \Delta t$, and (c) $t = 145.5 \Delta t$. The
vertical bars show the exciton condensate density $n(x,y)$. It is seen that
a traveling exciton pattern does not emerge for a large eccentricity $%
\protect\varepsilon$, and a turbulent condensate \protect\cite%
{Berman:12,Berman:12b} is formed instead. }
\label{fig:eps08}
\end{figure}

To more fully characterize the pattern rotation, we calculated the time
dependence of the probabilities $|c_m(t)|^2$ of having the specific angular
momentum $m$. Here, $c_m(t)$ are the normalized components the expansion of
the exciton condensate wave function over the basis functions with a fixed
momentum $m$ in the center-of-mass frame, 
\begin{equation}
c_m(t) = \frac{\displaystyle \int d \varphi_{\bm{\rho}} d \rho \, \rho \Psi( %
\bm{r},t) \Phi_m^*(\varphi_{\bm{\rho}}) } { \displaystyle \left( \int d
\varphi_{\bm{\rho}} d \rho \, \rho |\Psi( \bm{r},t) |^2 \right)^{1/2}} \, ,
\label{eq:cm}
\end{equation}
where in Eq.\ (\ref{eq:cm}) one has $\bm{r} = \bm{r}_{cm} + \bm{\rho}$, $%
\bm{r}_{cm}$ is defined in Eq.\ (\ref{eq:rcm}), $\Phi_m(\varphi_{\bm{\rho}})
= {\frac{1 }{\sqrt{2\pi}}} e^{i m \varphi_{\bm{\rho}}}$ is the eigenfunction
of a system with a given angular momentum $m$. The components (\ref{eq:cm})
are normalized so that $M(t) = \sum_m |c_m(t)|^2 m$.  Fig.\ \ref%
{fig:momentum}b shows the results of the calculations for $|c_m(t)|^2$  for
the pattern in Fig.\ \ref{fig:ring}a-c.  Specifically, the components with
the angular momenta $m=1$, 5, and 9 are plotted. It is seen that the
probabilities $|c_m(t)|^2$ strongly fluctuate with time. These fluctuations
lead to the fluctuations of the angular momentum $M(t)$ of the pattern seen
in Fig.\ \ref{fig:momentum}a. The fluctuations are caused by the energy
redistribution between the components with different $m$ due to the
nonlinear interactions in the condensate as well as due to the external pump.

Figure \ref{fig:momentum}c shows the time-averaged probabilities 
\begin{equation}
\langle|c_m|^2\rangle = {\frac{1}{t_2-t_1}} \int_{t_1}^{t_2} dt\, |c_m(t)|^2
\label{eq:maver}
\end{equation}
for the components with $-20\leq m\leq 20$. The probabilities in Eq. (\ref%
{eq:maver}) are averaged over the time interval between $t_1=140 \Delta t$
and $t_2=200\Delta t$. It is seen in Fig.\ \ref{fig:momentum}c that despite
the averaged angular momentum of the system oscillates around the mean value 
$M\approx 0.8$, the rotational motion of the pattern is more complex and
includes the components with the momenta reaching $|m| \sim 10$. Moreover,
both positive and negative $m$ contribute to the total angular momentum
therefore, rotations in both directions are present in the ring pattern at
the same time. If a vortex-like rotation were realized in the system, the
condensate density inevitably turned to zero at the center of the vortex, as
is described above. Additionally, the vortices with $|m|>1$ are dynamically
unstable and tend to decay to elementary vortices with $m=1$ \cite%
{Khalatnikov:65}. However, in our case, the condensate density does not turn
to zero at any point inside the pattern for both hill-like and ring-like
structures (see Fig.\ \ref{fig:d}) and its spontaneous rotation includes the
components with $|m|>1$ as is demonstrated in Fig. \ref{fig:momentum}c.
Therefore, the solid-body-like rotation is realized in the case of the
dipolar exciton condensate patterns. We emphasize that this average
solid-body-like rotation is a superposition of a few interacting rotating
flows of the condensate in the pattern, which interference results in the
fluctuations of the condensate density of the pattern, as is clearly seen in
Fig.\ \ref{fig:ring} above.

\subsection{Effects of the eccentricity and pumping rate}

We also studied the effects of the eccentricity $\varepsilon$ of the
trapping potential on the condensate dynamics. It was found that for
symmetric or slightly asymmetric traps with the eccentricity  $\varepsilon <
0.5$ the oscillating soliton patterns were formed similarly to those
described in the previous sections. However, for relatively large
eccentricities $\varepsilon \geq 0.5$ formation of the oscillating patterns
was not observed. The result of the simulations for $\varepsilon = 0.8$ is
shown in Fig.\ \ref{fig:eps08}. The absence of the oscillating patterns for
large $\varepsilon$ can be understood as follows. For a large eccentricity,
the width of the spatial domain in the $y$-direction energetically
accessible for the condensate in the trapping potential becomes comparable
with or smaller than the radius of the pattern $\rho_0$ and the pattern  is
squeezed in the $y-$direction by the trap. Due to that, the pattern becomes
unstable and  the exciton cloud filling all energetically accessible spatial
domain in the trap is formed, as is seen in Fig.\ \ref{fig:eps08}.  In this
case, the observed condensate dynamics are similar to a turbulent exciton
condensate described in Refs.\ \onlinecite{Berman:12} and %
\onlinecite{Berman:12b}, where nonlinear fluctuations of the cloud density
are observed rather than the formation of a traveling, spatially localized
soliton-like condensate structure.

\begin{figure}[t]
\begin{center}\vspace{-1.cm}
\includegraphics[width=7cm]{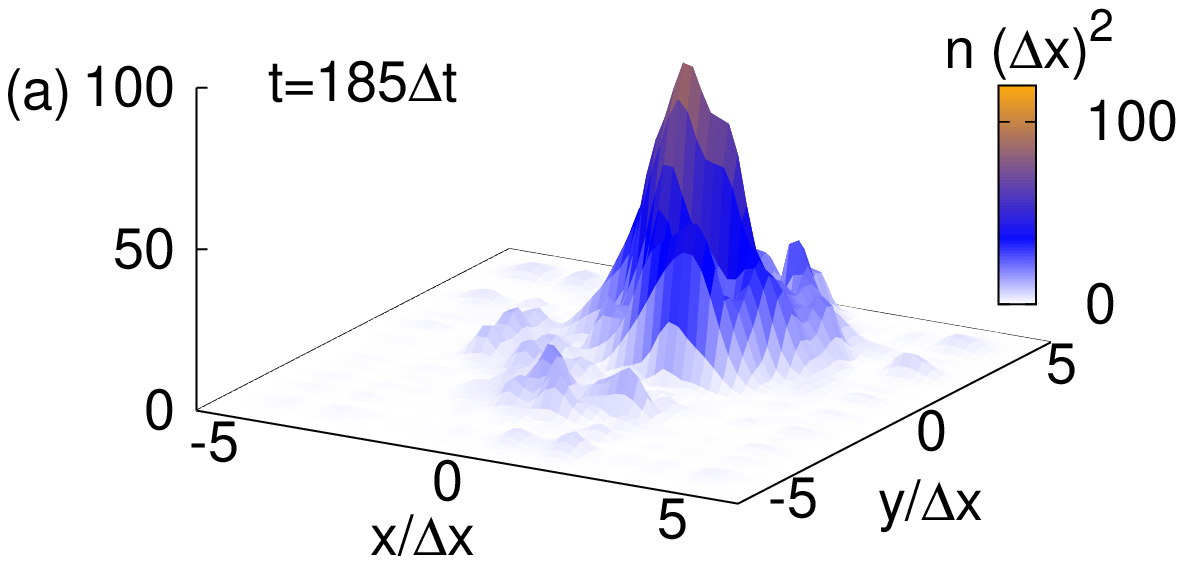}\\\vspace{-1.cm}
\includegraphics[width=7cm]{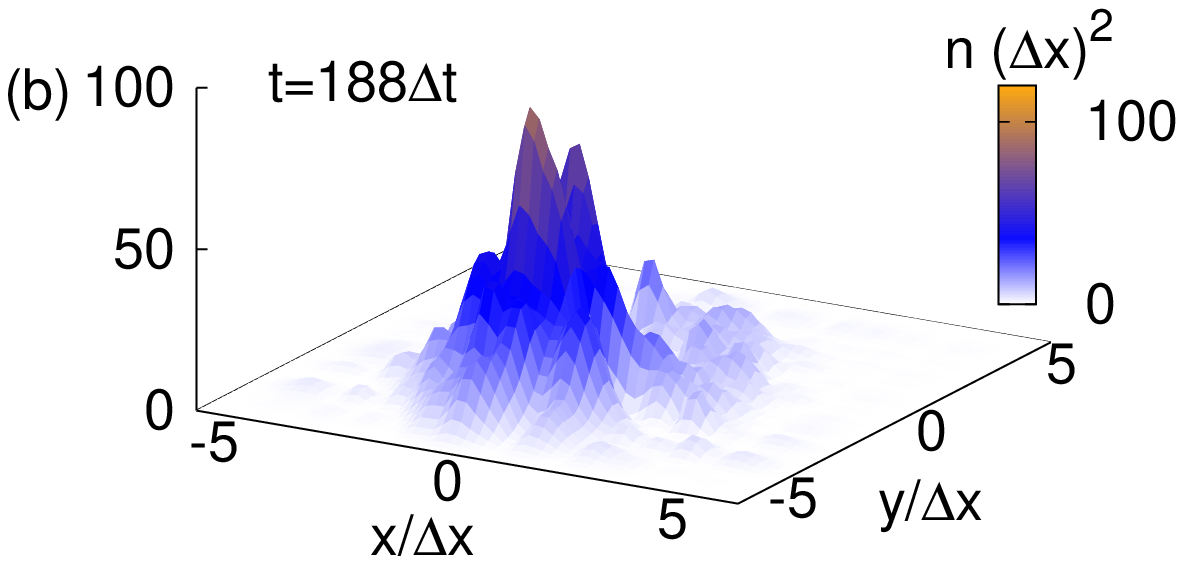}\\\vspace{-1.cm}
\includegraphics[width=7cm]{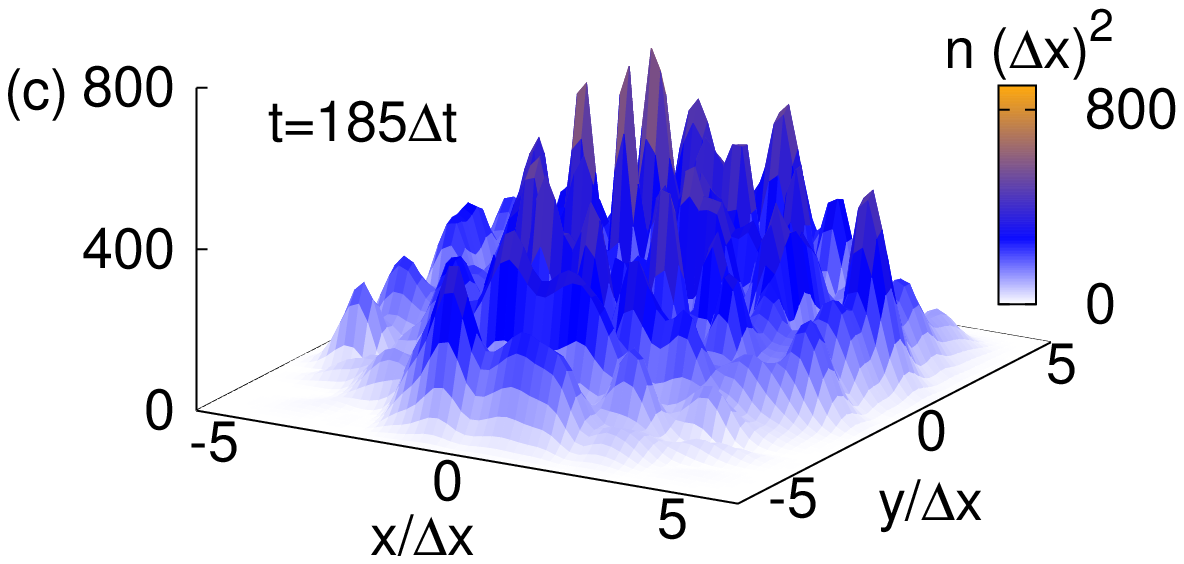}\\\vspace{-1.cm}
\includegraphics[width=7cm]{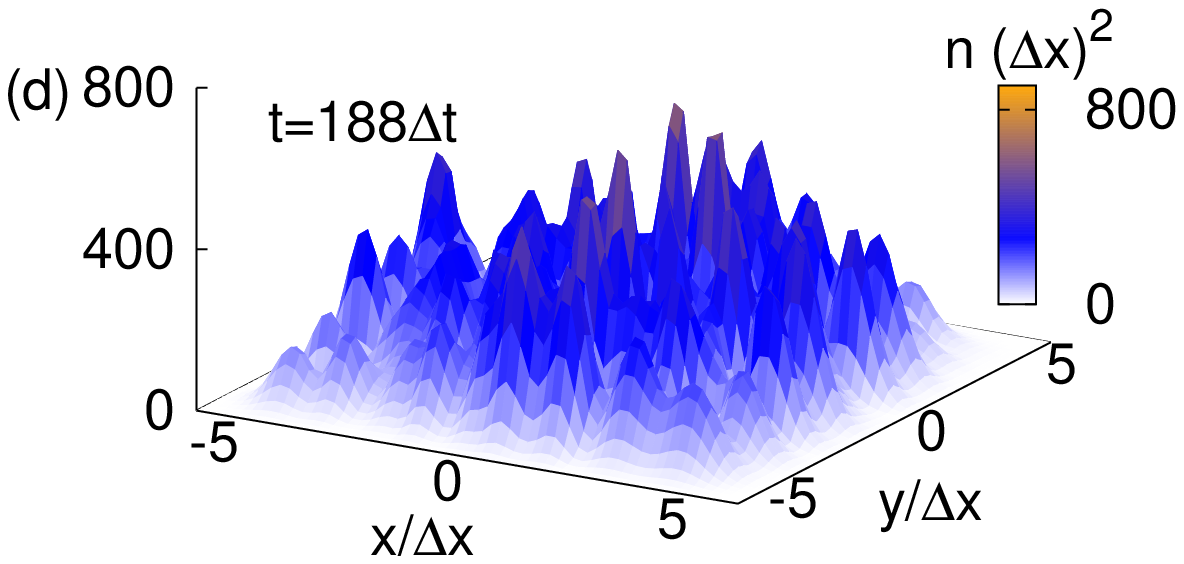}\\\vspace{-1.cm}
\end{center}
\par
\caption{ (Color online) 
Evolution of the condensate dynamics from  a traveling soliton (a,b) to a turbulent condensate (c,d) 
with the increase of the particle creation rate from $R_0=0.15 \omega_0$ to 
$R_0=0.25 \omega_0$. The eccentricity of the trap is $\varepsilon=0$,
time $t$ is shown on the plot. 
}
\label{fig:smallr}
\end{figure}

{
It was also found that the soliton-like condensate is not formed even for $\epsilon \ll 1$  if the
particle creation rate $R_0$ is higher than a certain value. In the latter case, a turbulent condensate was observed in the simulations, in agreement with the previous works Refs. \onlinecite{Berman:12,Berman:12b}. In this turbulent state, a relatively large fluctuating cloud is formed at the center of the trap.
The transition from a soliton-shaped to turbulent condensate 
with the increase of the particle creation rate $R_0$ is shown in Fig.~\ref{fig:smallr}.
}

\subsection{Comparison of our findings with the existing results}

The main object to which our findings can be applied is a non-equilibrium
solid-state condensate, e.g. a condensate of excitons, bound electron-hole
pairs, in semiconductor heterostructure. Since the lifetime of excitons in a
single semiconductor QW is relatively short ($\sim$ a few ns) \cite%
{Feldmann:87}, the experiments with exciton condensates are mostly focused
on so-called dipolar (or, indirect) excitons in coupled QWs, where the positively
charged holes and negatively charged electrons are located in different QWs,
which are separated by a nanometer-wide semiconductor or dielectric barrier 
\cite{Lozovik:75a,Negoita:99,Butov:94}.  The dipolar excitons formed by
spatially separated charges in coupled QWs are characterized by the
relatively long exciton lifetime compared to excitons in a single QW due to
small electron-hole recombination rates suppressed by the barrier between
QWs. It is worth noting that various experiments have revealed rich
collective dynamics of dipolar excitons in coupled QWs and have demonstrated
the existence of different  phases in an electron-hole bilayer system. The
experimental works on excitonic phases in coupled QWs were reviewed in Ref.~%
\onlinecite{Butov:07}. The experimental progress toward probing the ground
state of an electron-hole bilayer by low-temperature transport was reviewed
in Ref.~\onlinecite{Das:11}. The recent progress in the theoretical and
experimental developments in the studies of a dipolar exciton condensate in
coupled QWs was reviewed in Ref.\ \onlinecite{Snoke:13a}.

Owing to nonzero average electric dipole moment of dipolar excitons, their
interactions are long-ranged and decrease with the distance as $\propto
r^{-3}$. The presence of the long-range interactions can significantly
modify the condensate dynamics under certain regimes \cite{Lozovik:78}. In
particular, in spatially homogenous two-dimensional systems the logarithmic
corrections to the chemical potential of the system become essential \cite%
{Lozovik:78}. However, it was demonstrated \cite{Berman:12,Berman:12b} that
under usual experiment conditions for the dipolar exciton condensation in
coupled QWs, the most significant contribution to the exciton-exciton
dynamics arises from short-range scattering. In effect, the dynamics of the
dipolar exciton condensate can effectively be described as the local one,
where, however, the effective interaction strength for the condensate
particles pairwise interactions is a function of the chemical potential in
the system. This quasi-local model has recently been applied to the studied
of the nonlinear evolution of dipolar exciton condensates in a
radially-symmetric trap \cite{Berman:12,Berman:12b}.

The observed self-organization of the dipolar exciton condensates may be
similar to the dissipative soliton formation known for open nonlinear
optical systems, lasers, and chemical systems \cite{Akhmediev:05}. Solitons
dynamics described by one-dimensional conservative Gross-Pitaevskii equation
without the source and decay terms (also known as nonlinear Schr{\"o}dinger
equation) was studied in details for the case where the trapping potential
is absent (see Refs.\ \onlinecite{Novikov:84} and \onlinecite{Dubrovin:01}
for review). 
However, in our case, the system is two-dimensional that makes inapplicable
the general analytical approaches developed for the integrable
one-dimensional nonlinear systems \cite{Dubrovin:01}. An extended
dimensionality of the systems compared to Refs.\ %
\onlinecite{Novikov:84,Dubrovin:01} results in the presence of additional
degrees of freedom, in particular, in the possibility for the soliton to
rotate.

It is also worth noting that the soliton propagation  has recently been
observed in Refs.\ \onlinecite{Amo:11,Sich:12,Christmann:14} for a
condensate of polaritons, which are a quantum superposition of the QW
excitons and cavity photons \cite{Carusotto:13}.  In Ref.\ %
\onlinecite{Amo:11} the solitons propagate as a perturbation of the
steady-state condensate density.  In Refs.\ \onlinecite{Sich:12} and %
\onlinecite{Christmann:14} the solitons were created by a tightly focused
``writing'' laser beam  and then, freely expanded or traveled in the
microcavity. In our case, the exciton condensate \textit{spontaneously}
forms a traveling soliton under spatially homogenous pumping that is, inside
the pumping spot. Formation of inhomogeneous polariton condensate patterns
have also been considered in Refs.\ \onlinecite{Fernandez:13} and %
\onlinecite{Christmann:12}. In those works, the inhomogeneities in the
condensate density were caused by the interactions of polaritons with a
cloud of non-condensed excitons and with structural defects in the cavity.
The ring pattern formation was earlier demonstrated for the dipolar excitons
in coupled QWs and multiple QWs in previous experimental and theoretical
works \cite{Butov:02a,Snoke:02a,Butov:04,Butov:07,Fluegel:11}.  In spatially
resolved measurements, the photoluminescence of the dipolar exciton pattern was
seen as a central laser excitation spot surrounded by a sharp ring-shaped
region with a diameter much larger than that for the spot and with large
dark regions separating the two. In the coupled QWs experiments the voltage
was applied perpendicular to the wells, causing the bands to tilt and the
electrons and holes to separate. Both the lifetime and the energy of dipolar
excitons with spatially separated electrons and holes in different QWs can
be tuned by the electric field. The lifetime of quasiparticles in those
systems was increased up to microseconds due to the charge separation in
different QWs \cite{Voros:05}. In the cases mentioned above the formation of
the external rings was caused by the recombination of in-plane spatially
separated electrons and holes. The electrically injected carriers 
 created by the external gate voltage
recombine with optically injected carriers of the opposite sign \cite%
{Butov:04,Snoke:03,Rapaport:04}. In the framework of the microscopic
approach formulated in terms of coupled nonlinear equations for the
diffusion, thermalization and optical decay of the particles, the formation
of the inner ring was explained by the fact that in the optically-pumped
area the exciton temperature is much larger than the lattice temperature. As
a result, the recombination decay of excitons is suppressed, but while these
excitons diffuse out of the optically-pumped area, they cool down and
eventually recombine resulting in a local increase of the photoluminescence
signal \cite{Ivanov:06}. In other words, those experiments did not involve
Bose-Einstein condensation of excitons. Formation of the stationary exciton
condensate cloud with the size comparable with the size of the laser pumping
spot was treated in the framework of the Thomas-Fermi approximation in
Refs.\ \onlinecite{Voros:06} and \onlinecite{Berman:06}. Ring polariton
patterns were also excited by an elliptic ring laser spot in experiments
Ref.\ \onlinecite{Manni:11}. In contrast to all mentioned cases, the traveling
soliton-like condensates  reported in this paper have a characteristic size
smaller than the excitation spot  and are not caused by  the kinetics of the
free carriers or by the interactions with the  structural inhomogeneities.
We infer that the soliton formation is caused by the interplay of the
Bose-stimulated growth of the condensate that tends to increase the
condensate particle density and the exciton-exciton repulsion that leads
to the particle spreading in the system.

It is known that the condensation in infinite two-dimensional systems is
impossible due to  the phase fluctuations that destroy the long-range order
in the system \cite{Fisher:88}. To achieve the exciton condensation in
quasi-two-dimensional QW structures and to reach high enough exciton
densities, at which the system undergoes the BEC transition, most of
experiments were conducted with excitons localized in an external trapping
potential. A few techniques were utilized to create the exciton traps,
including the application of non-uninform mechanical stress to a
semiconductor sample \cite{Negoita:99} or electrostatic traps \cite%
{Gartner:07,High:12}. In these cases, the shape of the exciton trap varied
from a radially-symmetric one to a lozenge-shaped, elongated trap \cite%
{Negoita:99,Gartner:07}.

\section{Conclusions}

\label{sec:conclusion} To summarize, a trapped non-equilibrium condensate
can exhibit Bose-stimulated self-organization under the conditions where it
is driven by a homogenous external sources in a range of frequencies. This
self-organization results in formation of spatially-localized structures
such as solitary condensate waves -- humps or rings. These solitons
oscillate in a trap in a manner similar to a classical particle and also can
spontaneously rotate. If the asymmetry of the trap exceeds $\varepsilon \geq
0.5$ a conventional fluctuating elongated dipolar exciton cloud is formed
instead of a solitary wave.
{ The dynamics of the condensate  also
depends on the  pumping rate. It was found that at increased particle creation rates,
the soliton-like condensate dynamics is changed to a turbulent regime. 
The transition to turbulence at high  rates can be attributed to 
the excitation of multiple interacting degrees of freedom in the system due to 
nonlinearity. Turbulent exciton condensates have earlier been 
considered in Refs.\ \onlinecite{Berman:12,Berman:12b}.
} 

The spontaneous rotation of the soliton condensate observed in our simulations
can be detected in experiments via polarization of the emitted light due to
the electron-hole recombination photoluminescence \cite{Lagoudakis:08,Manni:11}. Recently,
polarized light has been applied to the multiplexing information transfer
through an optical waveguide \cite{Barrera:06,Wang:12}. The future work on
the self-organized dipolar exciton condensate might elucidate the ways to
control the soliton rotation, to be utilized as a source of coherent
polarized light with a given angular momentum $m$. It is of interest for
prospective applications in the above-mentioned information technology to
investigate how the self-organized condensate reacts on the pulsed or
modulated pumping, to create a controllable source of coherent polarized
light signals. Self-organized soliton condensate patterns can also
potentially be applicable for the information transfer in exciton-based
integrated circuits \cite{High:08}.

\begin{acknowledgments}
The authors are grateful to Dr. K. Ziegler for the discussion of the results.
The authors are also grateful to the Center for Theoretical Physics at
New York City College of Technology of the City University of New York for providing
computational resources.  The authors 
gratefully acknowledge support from the ARO, grant \#64775-PH-REP.
G.V.K. gratefully
acknowledges support from the PSC--CUNY award
\#67143-0045.
\end{acknowledgments}

%

\end{document}